\documentclass[11pt]{article}
\pdfoutput=1 
\usepackage[utf8]{inputenc}

\usepackage{amstext,amsmath,amssymb,amsfonts,bbm}
\usepackage{hyperref}
\usepackage{authblk}
\usepackage{caption} 
\usepackage{epsfig} 
\usepackage{color} 
\usepackage{graphicx} 
\usepackage{braket} 
\usepackage{slashed} 
\usepackage{dsfont} 
\usepackage{amsthm}  
\usepackage{physics}
\usepackage{soul}

\usepackage{cite}

\allowdisplaybreaks[4]

\usepackage{psfrag}
\usepackage{tikz} 
\usetikzlibrary{arrows}
\usetikzlibrary{plotmarks}
\usetikzlibrary{patterns}
\usetikzlibrary{external}
\usepackage{pgfplots}
\pgfplotsset{compat=1.9}
\usetikzlibrary{patterns}
\usetikzlibrary{calc}
\usepackage[compat=1.1.0]{tikz-feynman}
\usetikzlibrary{automata,positioning}

\usepackage{float}  
\usepackage{subfig}
\usepackage[lmargin=50pt,rmargin=60pt,tmargin=60pt,bmargin=65pt]{geometry}

\newcommand{\be}{\begin{equation}}
\newcommand{\ee}{\end{equation}} 

\newcommand{\f}{\frac}

\let\a=\alpha \let\b=\beta  \let\g=\gamma  \let\d=\delta
        \let\l=\lambda
               
\let\s=\sigma    \let\vph=\varphi  
    
\let\G=\Gamma     \let\X=F
         
  \let\eps=\epsilon

\newcommand{\bbeta}{\bar{\beta}}

\newcommand{\qb}{\bar{q}}

\newcommand{\psib}{\bar{\psi}}
\newcommand{\vphb}{\bar{\varphi}}

\newcommand{\cC}{\mathcal{C}}

\newcommand{\cO}{\mathcal{O}}

\newcommand{\tD}{\tilde{D}}
\newcommand{\hv}{\hat{v}}

\newcommand{\id}{\mathds{1}}

\allowdisplaybreaks[4]

\theoremstyle{remark}

\definecolor{orange}{rgb}{0.88,0.39,0.12} 
\definecolor{rouge}{rgb}{0.8, 0.0, 0.0}
\definecolor{vert}{rgb}{0.4, 0.69, 0.2}
\definecolor{bleu}{rgb}{0.19, 0.55, 0.91}
\definecolor{lavenderpurple}{rgb}{0.59, 0.48, 0.71}

\graphicspath{{figures/}}

\def\qb{{Q_{BRST}\,}}
\def\begs{\begin{split}}
\def\ends{\end{split}}
\def\eq#1{\eqref{#1}}

\begin{document}

\title{The Edge of Random Tensor Eigenvalues with Deviation}
\date{}

\author[1]{Nicolas Delporte\thanks{\rm\url{nicolas.delporte@oist.jp}}}
\author[2,3]{Naoki Sasakura\thanks{\rm\url{sasakura@yukawa.kyoto-u.ac.jp} }}

\affil[1]{\normalsize\it Okinawa Institute of Science and Technology Graduate University, 1919-1, Tancha, Onna, Kunigami District, Okinawa 904-0495, Japan.}

\affil[2]{\normalsize\it Yukawa Institute for Theoretical Physics, Kyoto University,\break
Kitashirakawa, Sakyo-ku, Kyoto 606-8502, Japan.}

\affil[3]{\normalsize\it CGPQI,  Yukawa Institute for Theoretical Physics, Kyoto University,\break
Kitashirakawa, Sakyo-ku, Kyoto 606-8502, Japan.}

\maketitle

\hrule\bigskip

\begin{abstract}

\noindent 
The largest eigenvalue of random tensors is an important feature of systems involving disorder, equivalent to the ground state energy of glassy systems or to the injective norm of quantum states. For symmetric Gaussian random tensors of order $3$ and of size $N$, in the presence of a Gaussian noise, continuing the work \cite{sasakura2023real}, we compute the genuine and signed eigenvalue distributions, using field theoretic methods at large $N$ combined with earlier rigorous results of \cite{auffinger2013random}. We characterize the behaviour of the edge of the two distributions as the variance of the noise increases. We find two critical values of the variance, the first of which corresponding to the emergence of an outlier from the main part of the spectrum and the second where this outlier merges with the corresponding largest eigenvalue and they both become complex. We support our claims with Monte Carlo simulations.
We believe that our results set the ground for a definition of pseudospectrum of random tensors based on Z-eigenvalues.

\end{abstract}

\hrule\bigskip

\section{Introduction}
While random matrices have become today a major tool in sciences, through a tremendous range of applicability \footnote{\cite{akemann2011oxford} is an excellent introductory reference to grasp the breadth of topics intimately related to random matrices, from chaos, RNA folding and QCD to random surfaces, knot theory and free probability.}, random tensors, omnipresent in classical and quantum networks, are still lacking the coherence that binds the matrices. Besides their use in data sciences, random tensors also serve as a discrete approach to quantum gravity \cite{ambjorn1991three,sasakura1991tensor,gross1992tensor} (a recent review is \cite{gurau2024quantum}) generating simplicial complexes from Feynman diagram expansions, attempting to generalize in higher dimensions the success of random matrices to produce and solve models of random surfaces with a large $N$ expansion that discriminates surfaces of different topology (a modern treatment is done in \cite{anninos2020notes}).

With tensors of size $N$ and of any order $p\geq 3$, an important milestone was achieved in \cite{gurau20111,Gurau_2011,Gurau_201102} that set up an action invariant under a tensor product $G^{\otimes p}$, for some compact Lie group $G$, leading to an analogous large $N$ expansion \footnote{The dependence in $N$ is coming from the choice of representation of the group $G$ or from the dimension of $G$.}, separating this time classes of graphs of different \emph{degree}. The leading order was shown to consist of \emph{melonic} graphs, enjoying a simplifying recursive structure that can be solved explicitly in that limit. This melonic limit turned out in fact to form a strong universality class competing with a two-dimensional Brownian sphere phase obtained with matrix-like interactions (and a more singular intermediate branched polymer phase) \cite{Lionni_2018}. Escaping the melonic universality class is still an open problem.

Matrices are often analysed through the lens of their spectral properties \cite{mehta2004random}, revealing aspects of universality and integrability.
For tensors, corresponding spectra are not as straightforward, since firstly there exist several definitions of eigenvalues \cite{qi2018tensor} and secondly, the number of the eigenvalues is typically exponential in the size of the tensor \cite{Cartwright_2013}. Other approaches involve studying the invariants under the tensor symmetry group \cite{gurau2020generalization}, although they grow factorially with the number of tensors contracted together \cite{geloun2020counting,klebanov2018tasi}. 
In fact, most tensor problems that generalize the linear systems of equations and inequalities of matrix problems (real eigenvalues, spectral norm, the best rank-1 approximation, etc.) are NP-hard \cite{hillar2013most}.
This offers another reason motivating the search of efficient approaches in the study of random tensors.

For a symmetric real tensor $C_{i_1\dots i_p}$ $(1\leq i_1,\dots, i_p\leq N)$ of order $p$, a Z-eigenpair \cite{QI20051302} is a real solution to the equations
\begin{gather}
\label{eq:eiva-def}
\sum_{1\leq i_2\dots i_p\leq N}C_{i_1\dots i_p}w_{i_2}\dots w_{i_p}=z w_{i_1}\,, \quad (1\leq i_1\leq N, z \in  \mathbb{R}, w \in  \mathbb{R}^N, \abs{w}=1),
\end{gather}
that is they form the zero set of
polynomials on $S_{N-1}$. When the tensor has random Gaussian entries, the polynomials are called Kostlan polynomials (see \cite{subag2023concentration} for the expectation value of the cardinality of the set, its variance at large $N$, as well as an exhaustive chronology of the results, originating with Kac \cite{kac1948average} and Rice \cite{rice1944mathematical}; \cite{breiding2019many} offers a different approach providing an exact counting of the average number of roots, from which asymptotic expressions at large $N$ are however difficult to extract). The problem is identical to the determination of the critical points of pure spherical $p$-spin models 
first solved with replicas \cite{crisanti1992spherical} and TAP equations \cite{crisanti1995thouless}.
In \cite{auffinger2013random} random matrix techniques allowed to obtain rigorously asymptotics and finite $N$
expressions of the expected number of critical points (its normalized logarithm is called in
the spin glass community the “complexity”) and the asymptotic expected distribution of critical points
of given finite index.
Fluctuations around the ground state were described in \cite{subag2017extremal} as following a Gumbel distribution. We refer to \cite{ros2022high} for a comprehensive overview of the history in understanding the critical points of high-dimensional random landscapes. The scaling of the largest real eigenvalue $z_{max}\sim \sqrt{N}$ was also obtained from concentration \cite{tomioka2014spectralnormrandomtensors} or combinatorial methods \cite{Evnin:2020ddw}. 

Offering a playground that connects complexity theory, glasses, neural networks, etc. \cite{parisi2023nobel}, many variations of the spherical $p$-spins have been studied:  mixed, Ising, with external field, etc.\footnote{See \cite{piccolo2023topological} for a concise guide to the recent literature.} The last case corresponds to a Hamiltonian of the form 
\begin{equation}
\label{eq:SphWithRandomField}
H= - \sum_{1\leq i_1\dots i_p\leq N} J_{i_1 \dots i_p}x_{i_1}\dots x_{i_p} - \sum_{i=1}^N h_i x_i ,\quad \abs{x}^2=N,\quad \expval{J_{i_1 \dots i_p}^2}=\frac{J^2}{pN^{p-1}}\,,
\end{equation}
where the external field $(h_i)_{1\leq i\leq N}$ can be deterministic or random, of norm or standard deviation $h$. It was shown \cite{fyodorov2015high,belius2022triviality} that for $h$ large enough, the number of critical points is trivial (there is only one maximum and one minimum) whereas after an explicit threshold, it becomes exponential. In the presence of a quadratic potential \cite{grela2022glass}, the density and distribution of stationary points grow more and more complex, as the deviation from a unit mass scales with $N^{-1/6}$ or $N^{-1/2}$. 
Taking a more general but deterministic potential for the spherical $p$-spin model, \cite{ros2019complex} gave a classification of the different types of landscapes (location and number) of critical points depending of the first and second derivative of the deterministic term.
The dynamics of such systems is known to display glassy behaviour \cite{Altieri2020}, but we will restrict ourselves to questions of equilibrium. 

Once the eigenpairs have been obtained, two crucial problems of practical importance are to determine if the eigensystem is stable in the presence of a small deformation and if it can be computed algorithmically. The deformed spectrum in the presence of noise is called \emph{pseudospectrum}. For a matrix $A$ and a deformation $\epsilon>0$, one definition of its pseudospectrum is:
\be
\sigma_\epsilon(A)=\{\lambda\in \mathbb{C}~\vert ~\norm{(A-\lambda I)^{-1}}\geq 1/\epsilon\}\,,
\ee
for a choice of norm $\norm{\cdot}$ \cite{trefethen2020spectra}. Equivalently, it is given by the set
\be
\sigma_\epsilon(A)=\{\lambda\in \mathbb{C}~\vert~\exists v\in \mathbb{C}^N, \abs{v}=1 \text{ such that } \abs{(A-\lambda I) v}\leq \epsilon\}\,.
\ee
The pseudospectrum of random matrices with independent entries with variance profile and deterministic diagonal deformation was studied recently in \cite{alt2024spectrum}. 
It is also a well-known issue that exact eigenvectors of matrices are uncomputable \cite{osinenko2020constructive}.
Given the diversity of eigenvalue and eigenvector definitions of tensors, such questions have to be reconsidered carefully. Recently, a corresponding notion of pseudospectra for tensors was developed, based on H-eigenvalues \cite{che2017pseudo,he2020pseudospectra} (see also \cite{lim2008spectrum} for a suggestion based on the smallest singular value of a tensor). Our work relates instead to the determination of the boundary of the pseudospectrum of large Gaussian tensors, for Z-eigenvalues.

Rescaling $z=1/\abs{v}^{p-2}, w=v/\abs{v}$ in \eqref{eq:eiva-def} and adding a Gaussian random deviation $\nu=(\nu_i)_{1\leq i\leq N}$ of variance $2\beta$ leads to the following eigenvector equation
\begin{gather}
\label{eq:eivec-def}
\sum_{1\leq i_2\dots i_p\leq N}C_{i_1\dots i_p}v_{i_2}\dots v_{i_p}= v_{i_1}+\nu_{i_1}\,, \quad (1\leq i_1\leq N, v \neq 0, v \in \mathbb{R}^N)\,.
\end{gather}
The deviation can be seen as the stochastic noise of a Langevin dynamics ($\beta$ corresponding to the temperature), or an error as one tries to solve the eigenvector equation. 
It bears resemblance to the case of an external potential, although the outcomes and the appropriate large $N$ behaviour turn out different. 
In our setting, we observe that as the variance of the noise increases, the number of eigenvalues remains exponential in $N$.

In the case of random matrices, it is well-known \cite{peche2014deformed} that outliers escape the limiting spectral distribution of a random matrix, the noise, in the presence of an added deterministic matrix, the signal, when the eigenvalues of the signal matrix are above a given threshold, corresponding to the BBP transition \cite{baik2005phase}. In other words, information about the signal can be recovered, when it is sufficiently strong compared to the noise. It was also shown \cite{baik2005phase} that the fluctuations of the largest eigenvalue differ from the Tracy-Widom law of the Wigner ensemble when the signal is detectable.
Here, we show that the second largest eigenvalue emerges from the main body of the spectrum and later merges with the corresponding trivial largest eigenvalue\footnote{We call ``trivial" the eigenvector $v$ (or the corresponding eigenvalue $\abs{v}^{2-p}$), the smallest in Euclidean norm eigenvector, since, as $\beta$ goes to zero, it continuously deforms to the ``trivial" solution to eq.~\eqref{eq:eivec-def}, $v=0$.}, in a sense becoming an outlier. 

Another example of a hard tensor problem is the spiked tensor model \cite{richard2014statistical}, asking for the detectability and recovery threshold of a given signal vector, hidden in a tensor to which a random tensor is added. 
In contrast with the matrix case that possesses an easy (with power iteration) and an impossible to recover the signal phase, separated by the BBP transition, it seems that the tensor case contains an intermediate ``hard” regime, above the information-theoretically possible threshold, where there is no polynomial time algorithm that recovers the signal \cite{arous2019landscape} (see however \cite{ouerfelli2023selective} for some promising methods based on tensor invariants). We emphasize that our approach takes the recovery problem from a different point of view, studying how the presence of a vector-like noise is making it difficult to retrieve the solutions to eigenvector equations beyond a certain threshold.

Focusing on tensors of order 3, one of the authors recovered the distribution of \cite{auffinger2013random} with a field theoretic approach at large \cite{sasakura2024signed} and finite $N$ \cite{Sasakura_2023}, reviving the earlier techniques that were approximating the absolute value of the determinant with the determinant (what below we call the signed distribution) or the use of replicas (see e.g. \cite{kurchan1991replica}). In \cite{sasakura2023real}, similar techniques were extended to the situation where the tensor has a non-zero background value \footnote{In the case where the background was a rank one tensor, the detectability threshold of \cite{arous2019landscape} was recovered.} and a Gaussian noise was introduced in the eigenvector equation, work that we will pursue. 
More recently, by matching the edge (signalling the largest eigenvalue) and the critical point of the signed distribution to those of the genuine spectral density, the case was made for some useful features that are contained in the signed spectrum, easier to compute exactly \cite{Kloos:2024hvy}. 

The aim of this program is to promote the use of techniques inspired from field theory in the study of random tensors that are suitable to understand large size phenomena, combined with rigorous methods from probability theory, and compare further the limits of what can be learned about the original problem from some approximations that nonetheless lead to simple and explicit formulas. 
The underlying method is to write an effective action for quantum fields, that symmetries and a correct decomposition of the degrees of freedom can simplify. Schwinger-Dyson equations for two-point and higher order correlators often reduce the apparent difficulty of the problem.

We plan our discussion as follows. We first start in Sec.~\ref{sec:signed} by deriving the exact signed eigenvalue distribution in the presence of disorder at finite $N$ using a quartic fermionic theory. The section~\ref{sec:genuine} uses a quartic action with a combination of fermions and bosons to rewrite the absolute value of the determinant, that we analyze in detail. In the large $N$ limit, we can factorize the transverse (noting that the transverse part is independent of the disorder) and parallel parts of the action. Writing the large $N$ Schwinger-Dyson equations together with a supersymmetry, we obtain the limiting expression of the genuine spectral distribution. In Sec.~\ref{sec:largest}, we compare the asymptotics of the edge of the spectrum of both the signed and genuine distributions, for the former using a Lefschetz thimble decomposition and for the latter, asymptotic expressions of the large $N$ action, giving the same as the signed case, extending thus the results of \cite{sasakura2023real}. We conclude briefly in Sec.~\ref{sec:conclusion} presenting some open questions. Four appendices close our paper. App.~\ref{appendix:num} containing details on our simulations and App.~\ref{appendix:interpretation} giving more numerical support for the behaviour of the edge of the genuine spectrum, using the finite $N$ formulas for the spectral density from \cite{auffinger2013random}. Relying on the BRST invariance of the signed distribution, App.~\ref{app:largebeta} explains that the total number of eigenvectors is still exponential in $N$, despite their apparent decrease as the variance $\beta$ of the noise grows. Finally, App.~\ref{app:newScaling} presents some numerical explorations, looking at the inner product of the first two eigenvectors with and without noise, and comparing the total number of eigenvalues with a different normalization.

\section{Signed distribution}
\label{sec:signed}
Let us quickly remind the setting of \cite{sasakura2023real}. 
For the simplicity of our discussions, we consider a tensor $C$ of order three ($p=3$).
The eigenvector distribution $\rho$ is given by
\footnote{Our convention is to sum over repeated indices.}
\begin{gather}
\label{eq:genuineDist}
\rho(v,C,\nu)= \sum_{i=1}^{n_{sol}} \delta^{(N)} (v-v^i) = \abs{\det M} \prod_{a=1}^N \delta(v_a - C_{abc}v_b v_c+\nu_a)\,,\\
M_{ab}=\delta_{ab}-2C_{abc}v_c\,,
\end{gather}
where $\nu_a$ is a Gaussian random vector of variance $2\b$ and $\{v^i\}_{1\leq i\leq n_{\text{sol}}}$ solve the eigenvector equation~\eqref{eq:eivec-def}. 
Additionally, we take the tensor $C$ to be a fully-symmetric Gaussian random variable with $\#C=N(N+1)(N+2)/6$ independent components such that the averaged distribution is
\begin{gather}
\label{eq:startingexpression}
\rho(v) = A^{-1}\int_{\mathbb{R}^{N}} \dd\nu\, e^{-\abs{\nu}^2/4\b} \int_{\mathbb{R}^{\#C}} \dd C \, e^{-\a \abs{C}^2} \rho(v,C,\nu)\,,\\
\abs{C}^2=\sum_{1\leq a, b, c\leq N} C_{abc}C_{abc}\,,\quad \dd C=\prod_{1\leq a\leq b\leq c\leq N} \dd C_{abc}\,,\quad \dd\nu=\prod_{1\leq i\leq N} \dd \nu_{i} \quad \alpha,\b>0,
\end{gather}
and the normalisation
\be
A=  \int_{\mathbb{R}^{N}} \dd\nu\, e^{-\abs{\nu}^2/4\b}\int_{\mathbb{R}^{\#C}} \dd C\, e^{-\a \abs{C}^2}\,.
\ee

Using the connection between \eqref{eq:eiva-def} and \eqref{eq:eivec-def}, the eigenvector distribution $\rho(v)$ is related to the eigenvalue distribution $\tilde\rho(z)$ by:
\begin{gather*}
\tilde\rho(z)\dd z=\rho(\abs{v}=1/z)\abs{v}^{N-1}S_{N-1}\dd \abs{v}\,,
\end{gather*}
with the spherical volume $S_{N-1}=2\pi^{N/2}/\Gamma(N/2)$, using also the invariance of the measure of the tensor under $O(N)$.

We recall some useful formulas (e.g. \cite{giardina2007course}),
\be 
\int_{-\infty}^{+\infty}\dd\lambda\, e^{i \lambda x} = 2 \pi \delta(x)\,,\quad \int_{\mathbb{R}^N} \dd \phi \, e^{-\phi^T M \phi}=\f{1}{\sqrt{\det (M/\pi)}}, 
\ee
where $x$ is real, and $M$ is an $N$-dimensional positive definite matrix,
and 
\be 
\quad \int \dd \psib\dd \psi \, e^{\psib M \psi}=\det M,
\ee
where $\psi,\psib$ are $N$-dimensional Grassmann variables, and 
$M$ an $N$-dimensional matrix.
We will first compute the signed eigenvector distribution following the steps of \cite{sasakura2024signed} :
\begin{align}
\begin{split}
\rho_s(v)&=\expval{\sum_{i=1}^{n_{sol}} \text{sign}(\det M(v^i)) \d^{(N)}(v-v^i)}\\
&=A^{-1}\int \dd \nu\dd C  \exp(-\abs{\nu}^2/4\b-\a \abs{C}^2) \det M(v) \prod_{a=1}^N\d(v_a-C_{abc}v_bv_c+\nu_a)\\
&=A^{-1}(2\pi)^{-N}\int \dd C \dd \psi \dd \psib \dd \lambda\, e^S
\end{split}
\end{align}
where, after integrating over the deviation $\nu$, the action is 
\begin{gather}
\label{eq:signaction}
S= - \a \abs{C}^2  - \b \abs{\lambda}^2 + i \lambda_a(v_a - C_{abc}v_b v_c) + \psib_a (\d_{ab}- 2C_{abc}v_c) \psi_b\,.
\end{gather}

After the integration over $C$ that cancels the normalizing constant $A$, the action becomes
\be
 - \b \lambda^2 + i \lambda \cdot v + \psib\cdot \psi + \f{1}{\a}\left(\f{1}{6}\sum_s \left(\frac{i}{2} \l_{s_a} v_{s_b} v_{s_c} + \psib_{s_a} \psi_{s_b} v_{s_c} \right)\right)^2\,,
\ee 
where the last sum is over all permutations $(s_a,s_b,s_c)$ of the indices $(a,b,c)$, since the tensor $C$ is fully symmetric. 
Gathering the terms containing the Lagrange multiplier $\l$, we find 
\begin{gather}
S_\l=-\f{\abs{v}^4}{12\a}B_{ab}\l_a\l_b +i\l_a \Delta_a,\\
B_{ab}= \left(1+ \frac{12 \a\b}{\abs{v}^4}\right) \d_{ab} + 2  \frac{v_a v_b}{\abs{v}^2} ,\\
\Delta_a=v_a + \f{1}{3\a}(\psib_a \psi\cdot v \abs{v}^2 + \psib\cdot v \psi_a  \abs{v}^2 + \psib \cdot v \psi\cdot v v_a),
\end{gather}
that can again be integrated exactly. It is convenient to project the Grassmann variables on the unit vector $\hat{v}_a=v_a/ \abs{v}$: $\psi_\parallel = \psi\cdot \hat{v}$ and $\psi_\perp = \psi -\hat{v} \psi_\parallel$. It is then easy to write the inverse of the covariance matrix~$B$
\begin{align} 
\begin{split}
B^{-1}_{ab}&=\f{\abs{v}^4}{12 \a \b+ \abs{v}^4}\id_{\perp ab}+\f{\abs{v}^4}{12 \a \b+ 3\abs{v}^4}\ \id_{\parallel a b}\\
&=\f{\abs{v}^4}{12 \a \b+ \abs{v}^4}\d_{ab}-\f{2 \abs{v}^{6}}{(12 \a \b+ 3 \abs{v}^4) (12 \a \b+\abs{v}^4)}v_av_b\,.
\end{split}
\end{align} 
After integrating over $\lambda$, we obtain the effective action 
\begin{align}
\begin{split}
S_\psi=&-\f{1}{2}\ln (3\abs{v}^4+12\a\b) -\f{N-1}{2} \ln (\abs{v}^4 + 12\a\b)-\f{\a \abs{v}^2}{\abs{v}^4+4\a\b}\\
&+ \psib_\perp\cdot\psi_\perp - \f{\abs{v}^4-4\a\b}{\abs{v}^4+4\a\b}\psib_\parallel \psi_\parallel -  \f{8 \b \abs{v}^2}{\abs{v}^4+12\a\b}\psib_\perp\cdot\psi_\perp \psib_\parallel \psi_\parallel -\f{\abs{v}^2}{6\a}(\psib_\perp\cdot\psi_\perp)^2,
\end{split}
\end{align}
where the first two terms arise from the determinant of the covariance $B$. Note that the quartic term mixing parallel and transverse components is new compared to \cite{sasakura2024signed}.
Doing next the integral over the parallel components gives
\begin{gather}
\int \dd \psib_\parallel\dd \psi_\parallel\, e^{S_\psi}=-\left(\f{\abs{v}^4-4\a\b}{\abs{v}^4+4\a\b}+\f{8 \b \abs{v}^2}{\abs{v}^4+12\a\b}\psib_\perp\cdot\psi_\perp \right) e^{S_{\psi_\perp}(k=1)+S_0},\\
S_{\psi_\perp}(k)=k\,\psib_\perp\cdot\psi_\perp-\f{\abs{v}^2}{6\a}(\psib_\perp\cdot\psi_\perp)^2,\\
S_0 = -\f{1}{2}\ln (3\abs{v}^4+12\a\b) -\f{N-1}{2} \ln (\abs{v}^4 + 12\a\b)-\f{\a \abs{v}^2}{\abs{v}^4+4\a\b}
,
\label{eq:S0}
\end{gather}
and the integral over the transverse components can also be done as follows \cite{sasakura2024signed}
\begin{align}
\begin{split}
Z_\perp(k)=&\int \dd \psib_\perp\dd \psi_\perp\, e^{S_{\psi_\perp}(k)} \\
=&\sum_{n\geq 0} \f{1}{n!}\left(-\f{\abs{v}^2}{6\a}\right)^n\left(\f{\partial}{\partial k}\right)^{2n}\underbrace{\int \dd \psib_\perp\dd \psi_\perp \, e^{k\psib_\perp\cdot\psi_\perp}}_{=k^{N-1}}\\
=&\left(\f{3\a}{2\abs{v}^2}\right)^{1-N/2} k \, U\left(1-\f{N}{2},\f{3}{2},\frac{3\a k^2}{2\abs{v}^2}\right)\,,
\end{split}
\end{align}
with $U$ being a confluent hypergeometric function of the second kind.
Collecting all the integration constants, the signed spectral distribution writes as
\begin{gather}
\rho_s(v)=-\left(\frac{\sqrt{12 \pi \a}}{2 \pi}\right)^N\left(\frac{\abs{v}^4-4\a \b}{\abs{v}^4+4\a \b}+\frac{8\b \abs{v}^2}{\abs{v}^4+12\a\b}\partial_k\right)Z_\perp(k)\rvert_{k=1}e^{S_0}  , \\
Z_\perp(k)=k\left(\frac{3\a}{2\abs{v}^2}\right)^{1-N/2}U\left(1-\f{N}{2},\f{3}{2},\frac{3\a k^2}{2\abs{v}^2}\right)\,.
\end{gather}

\begin{figure}
\centering
\includegraphics[width=0.32\linewidth]{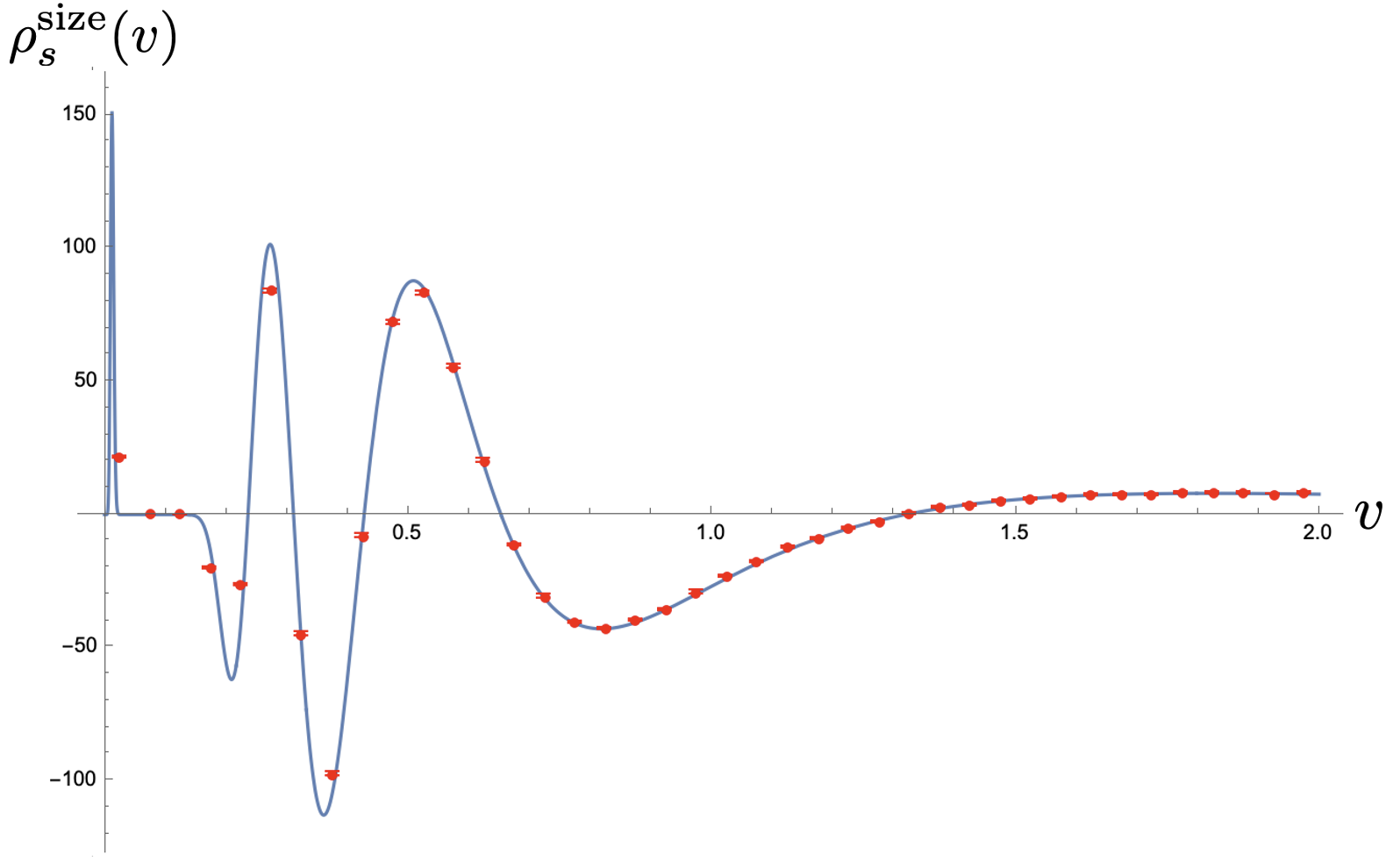} 
\includegraphics[width=0.32\linewidth]{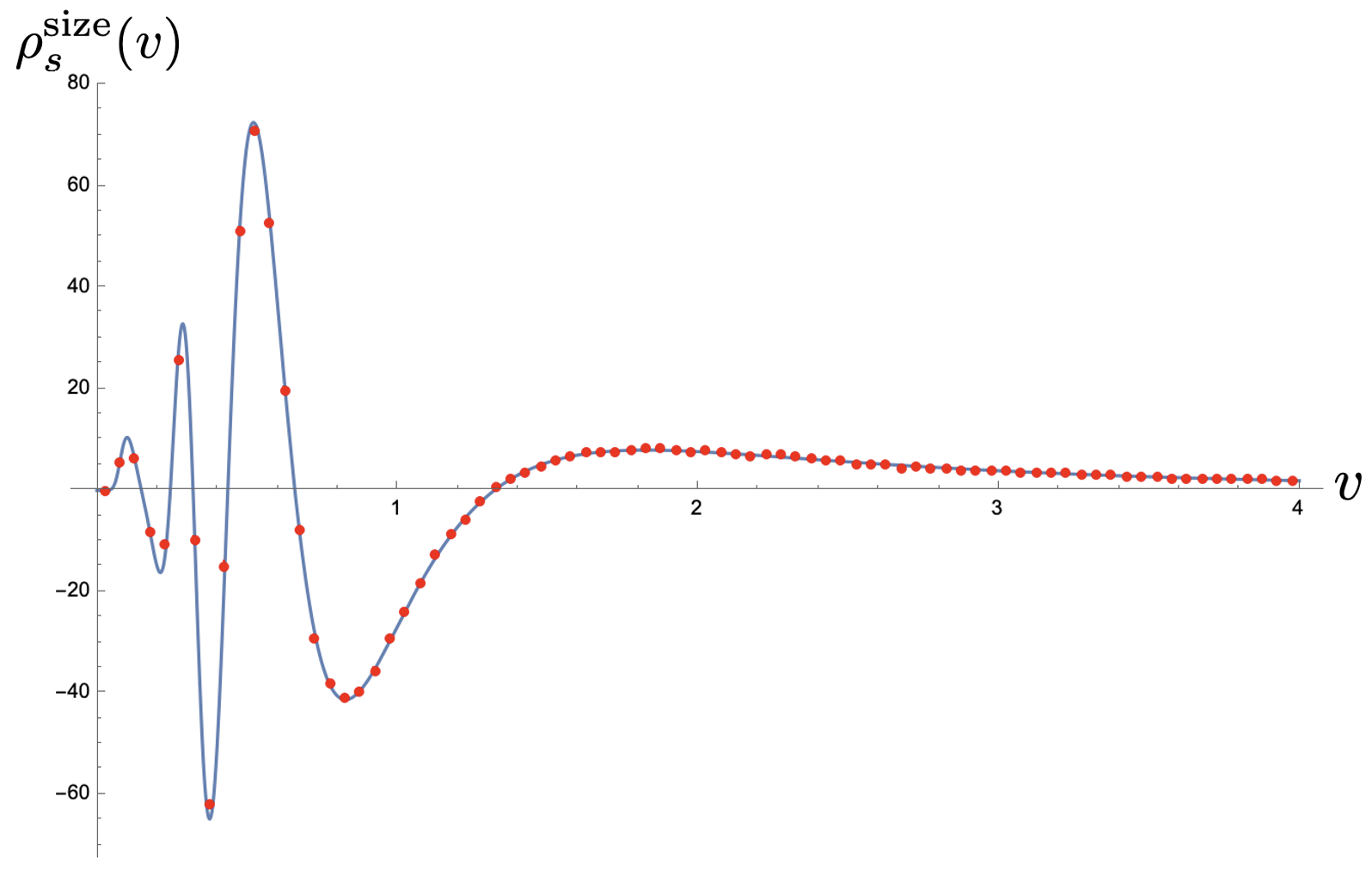} 
\includegraphics[width=0.32\linewidth]{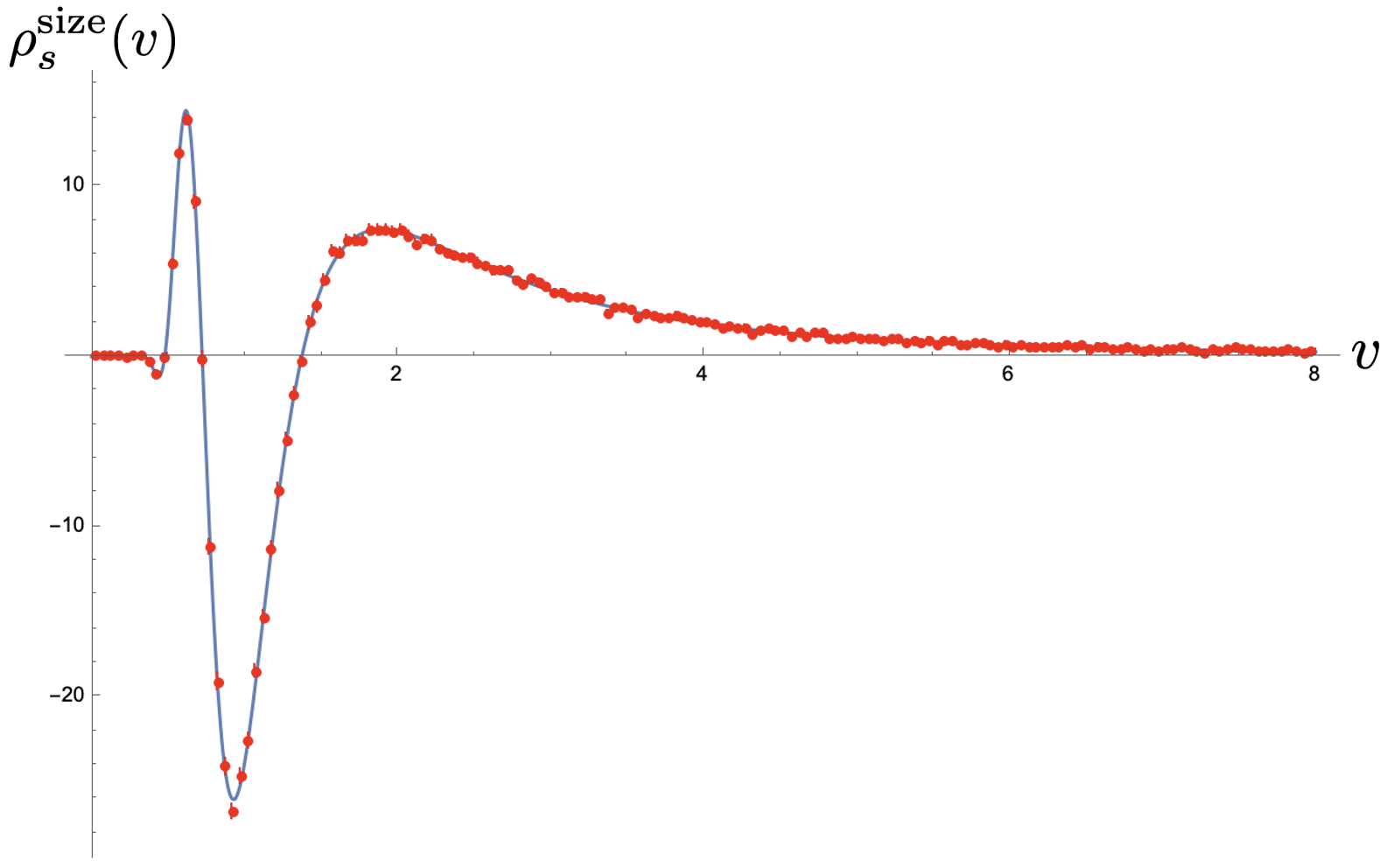} 
\caption{Comparison of the analytic expression (in continuous blue) of the signed eigenvalue distributions for $\beta=0.001/N^2,0.069/N^2,0.1$ (from left to right), with the numerical simulation (in dotted red, with associated numerical errors), for $N=12$ and 10000 iterations).}
 \label{fig:signed}
 \end{figure}

For the distribution of the absolute value $\abs{v}$, one looks at $\rho^\text{size}_s(|v|)=\rho_s(\abs{v})\abs{v}^{N-1}S_{N-1}$,
\be
\rho^\text{size}_s(|v|)=-\f{3 \a 2^{N/2}v^{2N-3}}{\G(N/2)}e^{S_0}\left(\frac{\abs{v}^4-4\a \b}{\abs{v}^4+4\a \b}+\frac{8\b \abs{v}^2}{\abs{v}^4+12\a\b}\partial_k\right)k\, U\left(1-\f{N}{2},\f{3}{2},\frac{3\a k^2}{2\abs{v}^2}\right)\bigg|_{k=1}\,.
\label{eq:rhosignfinal}
\ee
We compare in Fig.~\ref{fig:signed} the result of our analytic computation to Monte Carlo simulations (cf. Appendix~\ref{appendix:num} for the details on the simulations), for which we see excellent agreement. We also noted numerically that the integral of $\rho^\text{size}_s$ is independent of $\beta$ (see App.~\ref{app:largebeta} for the
explanation of this independence.).

\section{Genuine distribution}
\label{sec:genuine}
In this section, we will look at the genuine distribution eq.~\eqref{eq:genuineDist}, following \cite{Sasakura_2023}. 
We will first obtain the field theory corresponding to the genuine distribution in Section~\ref{sec:field}.
Then in Section~\ref{sec:compfield} we will compute the field theory partition function with approximations that we detail. We will
crosscheck our result with Monte Carlo simulations finding excellent agreement.

\subsection{Corresponding field theory and its symmetries}
\label{sec:field}
The absolute value of the determinant in \eqref{eq:genuineDist} can be written as
\be 
\abs{\det M}=\lim_{\eps \to 0^+} \f{\det (M^2 + \eps I)}{\sqrt{\det (M^2 + \eps I)}},
\ee 
and introducing the real scalar $(\phi_a)_{1\leq a\leq N}$ and Grassmann variables $(\psi_a, \bar{\psi}_a, \varphi_a, \bar{\varphi}_a)_{1\leq a\leq N}$, respectively for the denominator and the numerator, and after a Hubbard-Stratonovich procedure that decomposes the quadratic terms in $M$ into linear ones with the introduction of an extra Gaussian vector $(\sigma_a)_{1\leq a\leq N}$, we obtain the following integral
\be 
\rho (v,\a,\beta) = \lim_{\eps \to 0^+} \f{(-1)^N}{A (2\pi^2)^N}\int \dd C \dd \l \dd \phi \dd \s \dd \psib\dd \psi \dd  \vphb\dd \vph \, e^S,
\ee
\be
S=-\a \abs{C}^2 -\b \abs{\lambda}^2 + i \l_a(v_a -C_{abc} v_bv_c) -\abs{\s}^2 -2 i\s M\phi-\eps \abs{\phi}^2-\vphb\vph -\psib M\vph - \vphb M\psi +\eps \psib\psi\,,
\ee
where the quadratic term for the Lagrange multiplier $\l$ comes from the integration over the Gaussian noise. 

The integration over $C$ gives, like in \cite{Sasakura_2023}:
\begin{align} 
\begin{split}
S_{\setminus
C}&=-\f{\abs{v}^4}{12\a}\l_a B_{ab}\l_b-i\l_a(D_a+\tD_a)+E_1+E_2+E_3\,,\\
B_{ab}&=\d_{ab}\left(1+\f{12\a\b}{ \abs{v}^4}\right) + 2 \hv_a\hv_b=\id_{\perp ab}\left(1+\f{12\a\b}{ \abs{v}^4}\right)+ \id_{\parallel a b}\left(3+\f{12\a\b}{ \abs{v}^4}\right)\,,\\
D_a&=D_\parallel \hat{v}_a + D_{\perp a}\,,\\
\tD_a&=\tD_\parallel \hat{v}_a + \tD_{\perp a}\,,\\
E_1&=\f{1}{\a}\left(\f{1}{6}\sum_s(\psib_{s_a}\varphi_{s_b} v_{s_c} +\vphb_{s_a}\psi_{s_b} v_{s_c})\right)^2\,,\\
E_2&=-\f{4}{\a}\left(\f{1}{6}\sum_s v_{s_a} \sigma_{s_a}\phi_{s_a} \right)^2\,,\\
E_3&=\f{4 i}{\a}(v_a\psib_b\varphi_c+v_a\vphb_b\psi_c)\left(\f{1}{6}\sum_s v_{s_a} \sigma_{s_b}\phi_{s_c} \right)\,,
\end{split}
\label{eq:snotc}
\end{align} 
where $\hat{v}=v/\abs{v}$. The summation over $s$ encodes a sum over the symmetric permutations of all indices $a,b,c$. The parallel and transverse components to $v$ of $D,\tilde D$ are given by
\begin{align}
\begin{split}
D_\parallel&=\f{\abs{v}^3}{\a}(\bar{\psi}_\parallel\varphi_\parallel+\bar{\varphi}_\parallel\psi_\parallel)\,,\\
D_{\perp a}&=\f{\abs{v}^3}{3\a}(\bar{\psi}_{\perp a}\varphi_\parallel+\bar{\psi}_\parallel\varphi_{\perp a}+\bar{\varphi}_{\perp a}\psi_\parallel+\bar{\varphi}_\parallel\psi_{\perp a})\,,\\
\tD_\parallel&=\f{2 i \abs{v}^3}{\a}\s_\parallel\phi_\parallel\,,\\
\tD_{\perp a}&=\f{2 i \abs{v}^3}{3\a}(\s_\parallel\phi_{\perp a}+\s_{\perp a}\phi_\parallel)\,.
\end{split}
\end{align}
After integrating over $\l$ similarly as before, we obtain:
\begin{align}
\begs
S_{\setminus \{C,\lambda\}}=&-\f{\a \abs{v}^2}{\abs{v}^4+4 \a\b}+\f{2\a \abs{v}}{\abs{v}^4+4 \a\b}(D_\parallel+\tD_\parallel)\\
&-\left[\f{3\a}{\abs{v}^4+12\a\b}(D_\perp\cdot D_\perp +\tD_\perp\cdot\tD_\perp+2 D_\perp\cdot\tD_\perp)+\f{\a}{\abs{v}^4+4\a\b}(D_\parallel^2+\tD_\parallel^2+2 D_\parallel\tD_\parallel)\right]\,.
\ends
\end{align}
Putting them all together, the density can be rewritten as a zero-dimensional quartic field theory of bosons and fermions:
\begin{align}
\label{eq:FT}
\rho (v,\a,\beta) &= \lim_{\eps \to 0^+} \f{ 3^{(N-1)/2}\a^{N/2}}{\pi^{N/2}}\f{1}{(\abs{v}^4+4\a\b)^{1/2}(\abs{v}^4+12\a\b)^{(N-1)/2}}\exp(-\f{\a \abs{v}^2}{\abs{v}^4+4\a\b})Z_{FT}\,,
\\
&Z_{FT} = \frac{(-1)^N}{\pi^N} \int \dd B\dd F \exp(K_{\parallel}+\d V+S_{\perp})\,,
\label{eq:zftgenuine}
\end{align}
where $\int dB dF$ stands for the integration over all the bosonic and fermionic variables.
We have combined the terms quadratic in parallel components into $K_\parallel$:
\be
K_{\parallel}=
\begin{pmatrix}
\psib_\parallel&\vphb_\parallel
\end{pmatrix}
\begin{pmatrix}
\eps&\g_1\\
\g_1&-1
\end{pmatrix}
\begin{pmatrix}
\psi_\parallel\\
\vph_\parallel
\end{pmatrix}+
\begin{pmatrix}
\phi_\parallel&\s_\parallel
\end{pmatrix}
\begin{pmatrix}
-\eps&\g_1 i\\
\g_1 i&-1
\end{pmatrix}
\begin{pmatrix}
\phi_\parallel\\
\s_\parallel
\end{pmatrix}
\ee
and the mixing and quartic terms in the parallel components into $\delta V$:
\begin{align}
\label{eq:dV}
\begs
\d V=& -\g_2 \psib_\parallel\psi_\parallel\vphb_\parallel\vph_\parallel +2\g_2\left(i(\psib_\parallel\vph_\parallel+\vphb_\parallel\psi_\parallel) \s_\parallel\phi_\parallel - (\s_\parallel\phi_\parallel)^2\right)\\
&+2\g_3\left(i(\psib_\perp\vph_\parallel+\psib_\parallel\vph_\perp+\vphb_\perp\psi_\parallel+\vphb_\parallel\psi_\perp)\cdot(\s_\parallel\phi_\perp+\s_\perp\phi_\parallel) - (\s_\parallel\phi_\perp+\s_\perp\phi_\parallel)^2\right)\\
&+\g_3\left(\psib_\perp\cdot\vphb_\perp \psi_\parallel\vph_\parallel
+\psi_\perp\cdot\vph_\perp \psib_\parallel\vphb_\parallel
-\psib_\perp\cdot\psi_\perp \vphb_\parallel\vph_\parallel
-\psib_\perp\cdot\vph_\perp \psib_\parallel\vph_\parallel
-\vphb_\perp\cdot\vph_\perp \psib_\parallel\psi_\parallel
-\vphb_\perp\cdot\psi_\perp \vphb_\parallel\psi_\parallel \right),
\ends
\end{align}
introducing the constants: 
\be
\g_1=\f{\abs{v}^4-4\a\b}{\abs{v}^4+4\a\b}\,,\quad \g_2=\f{8\b \abs{v}^2}{\abs{v}^4+4\a\b}\,,\quad\g_3=\f{8\b \abs{v}^2}{\abs{v}^4+12\a\b}\,.
\ee
The terms involving only transverse components are collected in $S_\perp$ (notice that it is independent of $\b$), with the quadratic terms into $K_\perp$, and the quartic bosonic, fermionic and sharing bosonic and fermionic transverse components respectively into $V_B$, $V_F$ and $V_{BF}$:
\begin{align}
\begs
&S_\perp=K_{\perp}+V_F+V_B+V_{BF}\,,\\
&\ \ \ K_{\perp}=
\begin{pmatrix}
\psib_\perp&\vphb_\perp
\end{pmatrix}
\begin{pmatrix}
\eps&-1\\
-1&-1
\end{pmatrix}
\begin{pmatrix}
\psi_\perp\\
\vph_\perp
\end{pmatrix}+
\begin{pmatrix}
\phi_\perp&\s_\perp
\end{pmatrix}
\begin{pmatrix}
-\eps&- i\\
-i&-1
\end{pmatrix}
\begin{pmatrix}
\phi_\perp\\
\s_\perp
\end{pmatrix}\,,\\
&\ \ \ V_B=-\f{2\abs{v}^2}{3\a}\left(\s_\perp^2\phi_\perp^2 + (\s_\perp\cdot \phi_\perp)^2\right)\,,\\
&\ \ \ V_F=-\f{\abs{v}^2}{6\a}\left((\psib_\perp\cdot \vph_\perp)^2+(\vphb_\perp\cdot\psi_\perp)^2 + 2\psib_\perp\cdot \vphb_\perp\vph_\perp\cdot \psi_\perp+2 \psib_\perp\cdot \psi_\perp\vphb_\perp\cdot \vph_\perp\right)\,,\\
&\ \ \ V_{BF}=\f{2i\abs{v}^2}{3\a}\left(\psib_\perp\cdot \s_\perp\vph_\perp\cdot \phi_\perp+\vphb_\perp\cdot \s_\perp\psi_\perp\cdot \phi_\perp+\psib_\perp\cdot \phi_\perp\vph_\perp\cdot \s_\perp+\vphb_\perp\cdot \phi_\perp\psi_\perp\cdot \s_\perp\right)\,.
\ends
\label{eq:sperp}
\end{align}

\paragraph{Symmetries.}
Two symmetries are important to understand the large $N$ limit of the system: the global $O(1)\times O(N-1)$ symmetry,\footnote{Since the parallel and transverse sectors form inner products within the same sector.} leading to the factorization between the parallel and transverse components action in that limit, and the graded symmetry  exchanging the bosonic and the fermionic variables,
\begin{gather}
\d_1\psi_{ a}=-\phi_{ a}\,, \quad \d_1\phi_{a}=\f{1}{2}\psib_{ a}\,, \quad \d_1\vph_{a}=i\s_{ a}\,, \quad \d_1 \s_{ a} = \f{i}{2}\vphb_{ a}\,,\\
\d_2\psib_{ a}=\phi_{ a}\,, \quad \d_2\phi_{ a}=\f{1}{2}\psi_{ a}\,, \quad \d_2\vphb_{ a}=-i\s_{ a}\,, \quad \d_2 \s_{ a}
 = \f{i}{2}\vph_{ a}\,.
 \label{eq:susyrel}
\end{gather}
Being assumed unbroken \footnote{We have verified that those equalities hold for $N=1$. It is also not a proper supersymmetry since the number of bosons differ from the number of fermions and there is no space dependence \cite{parisi2006computing}.}, this symmetry implies the following relations between the expectation values of the fermionic and the bosonic fields, which will be assumed in Section~\ref{sec:compfield}:
\begin{align}
\label{eq:reltwopt}
\begs
&\expval{\psib_{\perp}\cdot\psi_{\perp}}=2\expval{\phi_{\perp}\cdot\phi_{\perp}},\quad \expval{\vphb_{\perp}\cdot\psi_{\perp}}=-2i\expval{\phi_{\perp}\cdot\s_{\perp}}, \\
&\expval{\vphb_{\perp}\cdot\vph_{\perp}}=-2\expval{\s_{\perp}\cdot\s_{\perp}}, \quad \expval{\psib_{\perp}\cdot\vph_{\perp}}=-2i\expval{\s_{\perp}\cdot\phi_{\perp}}.
\ends
\end{align}

\subsection{Computation}
\label{sec:compfield}
In this section, we will obtain the genuine distribution by explicitly
computing  the partition function \eqref{eq:zftgenuine} by an approximation, which is expected to be valid for large $N$. 
As shown in Section~\ref{sec:field}, the field theory can be regarded as an interacting system 
between the parallel and transverse sectors. 
For large $N$ the transverse sector dominates, and therefore we can neglect the parallel components to determine the dynamics of the transverse ones by their self-interaction. In a second step,
the dynamics of the parallel sector can be computed in the background of the transverse sector,
where the interaction between the parallel and transverse sectors is considered. 
More explicitly, in the large $N$ limit, the partition function factorizes into the form,
\be 
Z_{FT}\sim Z_{FT}^\perp \cdot Z_{FT}^\parallel(Q^*),
\label{eq:zftproduct}
\ee
where $Z_{FT}^\perp$ denotes the partition function determined solely by the transverse sector, and 
$Z_{FT}^\parallel (Q^*)$ denotes that of the parallel sector with $Q^*$ being the background of the 
transverse sector.

An important fact in our approximation is that $Z_{FT}^\perp$ does not depend on $\beta$, 
while the $\beta$-dependence appears in $Z_{FT}^\parallel(Q^*)$.
Therefore we can obtain $Z_{FT}^\perp$ using some former results without the noise.  
We will first review the Schwinger-Dyson (SD) method, which was employed in \cite{sasakuraSDE}. 
This determines  $Z_{FT}^\perp$ as well as the background $Q^*$  in the large $N$ limit. 
We will also use the exact formula given in \cite{auffinger2013random} for better approximations at finite $N$. 

\paragraph{Transverse sector.}
The transverse sector is defined by the partition function, 
\be
Z_{FT}^\perp = \frac{(-1)^{N-1}}{\pi^{N-1}} \int \dd B_\perp \dd F_\perp e^{S_{\perp}},
\ee
where $S_\perp$ is given in \eqref{eq:sperp}. Note that it does not contain $\beta$.
Neglecting the subleading terms in the potentials $V_F$ and $V_{BF}$ and assuming that the $O(N-1)$ symmetry remains unbroken in the large $N$ limit, we can write an effective transverse action in terms of the two-point correlators:
\begin{align}
S_\perp(Q)&=\f{N-1}{2}\left(\eps Q_{11}-Q_{22}-2 Q_{12} - x (Q_{12}^2+Q_{11}Q_{22})-\log(-\det Q)\right) +\cO(1)\,,
\label{eq:Sperp}
\end{align}
\be
Q=\f{1}{N-1}\begin{pmatrix}
\expval{\psib_{\perp}\cdot\psi_{\perp}}  &\expval{ \psib_{\perp}\cdot\vph_{\perp}}\\
\expval{\vphb_{\perp}\cdot\psi_{\perp}} &\expval{\vphb_{\perp}\cdot\vph_{\perp}}
\end{pmatrix} \,, \quad x=\f{(N-1)\abs{v}^2}{3\a}\,,
\ee
where we used that $Q_{12}=Q_{21}$,\footnote{This can be derived from \eqref{eq:reltwopt}.} and defined the variable $x$.
The logarithmic term arises as a Jacobian of the change of variables from the fermionic fields to the bilinear $Q$ (see \cite{zamponi2014mean}, section B4) \footnote{We introduced a sign since the saddle points $Q^*$ below 
satisfy $\det Q^*<0$.}.

Notice that, in deriving \eq{eq:Sperp}, we have used the supersymmetric relations \eq{eq:susyrel} to write all the two-point correlators in terms of fermionic ones. This can be done, because bosons and fermions have the same leading order action, so the saddle point is also supersymmetric, as was already observed in \cite{cavagna1999quenched}. Similar graded symmetries were used in earlier treatments of 
systems with rough landscapes, when approximating the absolute value of the determinant by the determinant itself. It is now 
understood that this supersymmetry holds if there is at most one step replica symmetry breaking, but it is broken if higher, when the number 
of metastable states is very sensitive to external perturbations, like in the Sherrington-Kirkpatrick model \cite{giardina2007course}.

\paragraph{Large-$N$ solution.}
Since the effective transverse action is independent of $\beta$ and contributions from the parallel sector are subleading in $N$, we can use the saddle point analysis of \cite{sasakuraSDE}. At large $N$, the leading order saddle point of $S_\perp$ that reproduces the free theory at $v=0$ is singular at the critical point $x=1/4$, and expanded at small $\epsilon$, is given by:
\begin{itemize}
    \item $0<x<1/4$:
        \begin{align}
        \begs
    \label{eq:Qsaddle1}
    &Q_{11}^* = \frac{-\sqrt{1-4x}+1}{2x\sqrt{1-4x} }+\cO(\eps)\,,\\
    &Q_{12}^*= \frac{1-\sqrt{1-4x}-4x}{2x\sqrt{1-4x}}+\cO(\eps)\,,\\
    &Q_{22}^* = \cO(\eps^2)\,,
    \ends
    \end{align}
    \item $x>1/4$:
        \begin{align}
        \begs
    \label{eq:Qsaddle2}
    &Q_{11}^* = \f{\sqrt{4x-1}}{2\sqrt{\eps}x} - \f{1}{2x}+\cO(\sqrt{\eps})\,,\\
    &Q_{12}^*=- \f{1}{2x} +\f{\sqrt{\eps }}{2x\sqrt{4x-1}}+\cO(\eps^{3/2})\,,\\
    &Q_{22}^* = -\f{\sqrt{\eps}\sqrt{4x-1}}{2x}+ \f{\eps}{2x}+\cO(\eps^{3/2})\,.
    \ends
    \end{align}
\end{itemize}
When $\eps\to 0^+$, the effective action evaluates to:
\begin{equation}
\label{eq:SperpVal}
S_{\perp}(Q^*) = 
\begin{cases}
    \frac{(N-1)}{4}\left(2+\log 16 +\frac{1-\sqrt{1-4 x}}{x}-4\log \frac{1-\sqrt{1-4x}}{x}\right)\,, & 0<x<1/4\,,\\
    \frac{(N-1)}{4x}\left(1+2x+2x\log x\right)\,, & x>1/4\,.
\end{cases}
\end{equation}
Thus the partition function in the large-$N$ is given by
\be
Z_{FT}^\perp\sim e^{S_{\perp}(Q^*)-(N-1)},
\ee
where the subtraction serves to ensure the normalization $Z_{FT}^\perp=1$ for  $|v|=0\ (x=0)$.

\paragraph{Parallel sector.}
Now we want to compute the remaining part of the partition function \eq{eq:zftgenuine}:
\be
Z_{FT}^\parallel(Q^*) =-\frac{1}{\pi} \int \dd B_\parallel \dd F_\parallel \, e^{K_\parallel+\delta V},
\ee
where all the transverse fields are replaced by the two-field correlators $Q^*$ obtained above. 
To deal with the parallel components, 
we can do a Hubbard-Stratonovich transformation with a Gaussian variable $g$ to split the quartic 
terms of the first line of eq.~\eqref{eq:dV} as
\be
\f{1}{\sqrt{\pi}}\int_{-\infty}^{+\infty} \dd g e^{-g^2 +\sqrt{2\g_2}(\psib_\parallel\vph_\parallel+\vphb_\parallel\psi_\parallel+2 i\s_\parallel\phi_\parallel) g}\,,
\ee
and integrate out the now Gaussian parallel components of the fermionic and bosonic fields (the last ones bringing each a factor $\sqrt{\pi}$).

For large $N$, the first term of the second line in $\d V$ is subleading, mixing parallel and transverse components of the bosonic and fermionic variables, as well as the first two of the third line. Therefore, the partition function factorizes since the resulting determinant factor after integrating the Gaussian parallel components is subleading in $N$ with respect to the purely transverse action \eqref{eq:sperp}, leading to
\be
\label{eq:zpara}
Z^{\parallel}_{FT}(Q^*)=\frac{-1}{\sqrt{\pi}}\int_{-\infty}^{+\infty} \dd g
\, e^{-g^2}\lim_{\eps\rightarrow 0^+}\left[\big(\eps-\g_3(N-1)Q^*_{22}\big)\big(1+\g_3(N-1)Q^*_{11}\big)+\left(\g_1-(N-1)\g_3Q^*_{12}+\sqrt{2\g_2}g\right)^2\right]^{1/2}\,.
\ee
This 
expression can easily be numerically integrated with Mathematica after manually taking the limit $\eps\to 0^+$.

\paragraph{Finite $N$ case.}
For finite $N$, we can take advantage of the exact formula given in 
\cite{auffinger2013random}.\footnote{See \cite{sasakuraSDE} for a field theoretical derivation.}
From \eq{eq:FT}, \eq{eq:zftproduct}, the independence of $Z_{FT}^\perp$ on $\beta$, 
and $Z_{FT}^\parallel(Q^*)=1$ for $\beta=0$, we find
\be
\label{eq:rhosizemoreexact}
\rho^{\rm size}(|v|,\alpha,\beta)\sim \left(\frac{|v|^4}{|v|^4+4 \alpha \beta}\right)^{\frac{1}{2}}
\left(\frac{|v|^4}{|v|^4+12 \alpha \beta}\right)^{\frac{N-1}{2}}\exp \left( -\frac{\alpha |v|^2}{|v|^4+4 \alpha \beta} 
+\frac{\alpha}{|v|^2} \right) Z_{FT}^\parallel(Q^*)\, \rho^{\rm size}(|v|,\alpha,\beta=0),
\ee 
where $\rho^{\rm size}(|v|,\alpha,\beta=0)$ is the genuine distribution without the noise. 
By properly changing the variables in \cite{auffinger2013random} and 
considering that their normalization of the Gaussian tensor corresponds to $\alpha=1/2$, 
we obtain 
\be
\rho^{\rm size}(|v|,\alpha=1/2,\beta=0)=2\sqrt{N}|v|^{-2}(p-1)^{(N-1)/2} \exp (-(p-2)/(4(p-1) |v|^2)) 
\rho_N(\tilde x), 
\ee 
where $\rho_N(\cdot)$ is given in Section 7.2 of \cite{auffinger2013random}, 
$\tilde x=-\sqrt{p/(2(p-1))}/(\sqrt{N} |v|)$, and $p=3$ (see Appendix~\ref{appendix:interpretation} for more explanation).
Note that the formula \eqref{eq:rhosizemoreexact} still depends on the SD approximation for $Q^*$, which is valid in the large-$N$ limit.

\begin{figure}
\centering
\includegraphics[width=0.32\linewidth]{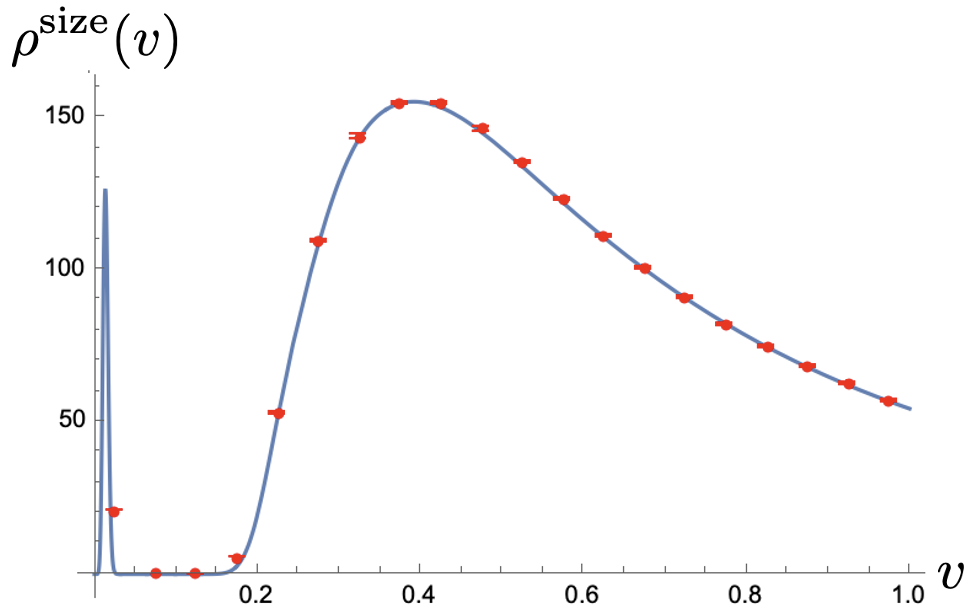}
\includegraphics[width=0.32\linewidth]{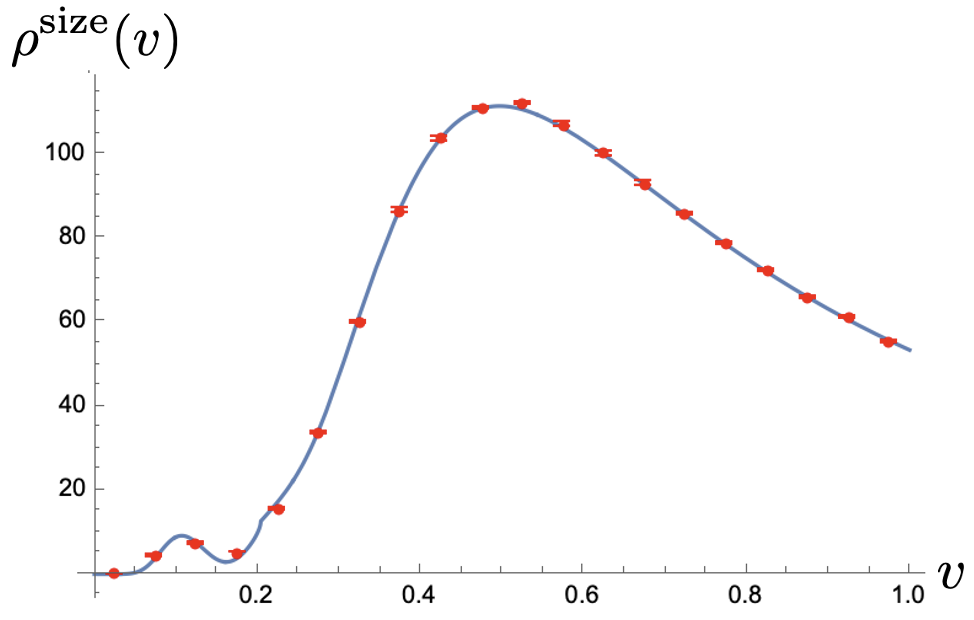}
\includegraphics[width=0.32\linewidth]{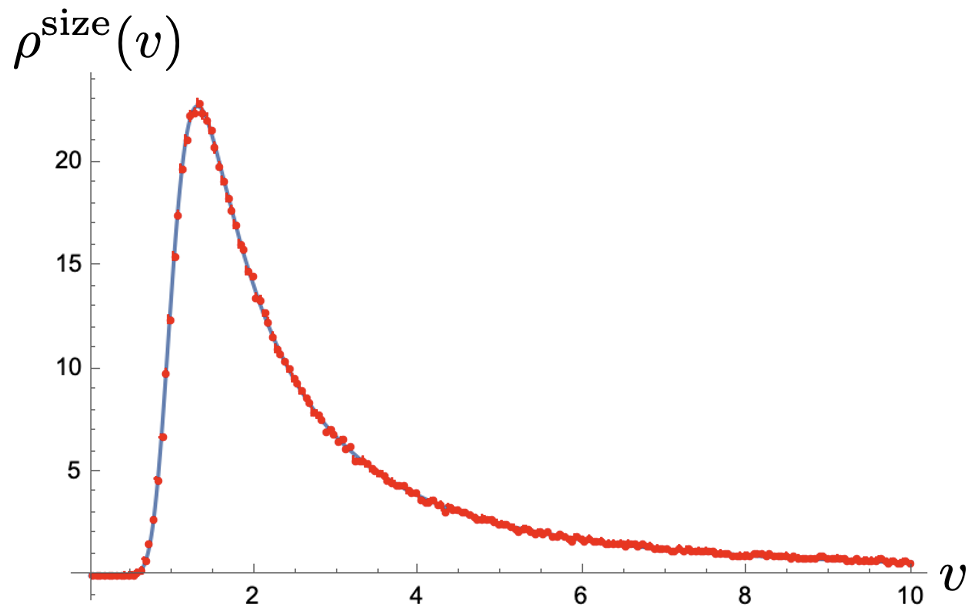}
 \caption{Genuine eigenvector distributions for $\beta=0.001/N^2,0.069/N^2,0.1$ (from left to right), the analytic expression (in continuous blue) and Monte Carlo simulation (in dotted red) with the associated numerical error, for $N=10$ and 10000 iterations.}
 \label{fig:numN10}
 \end{figure}

\paragraph{Summary.} We have crosschecked the expression \eq{eq:rhosizemoreexact} with Monte Carlo 
simulations (see Appendix~\ref{appendix:num} for the details). 
The agreement is shown in Fig.~\ref{fig:numN10} for different values of $\beta$. 
On the middle plot, we see the emergence of the outlier and on the rightmost
plot, its merging with the trivial eigenvector.
One also notices that as $\beta$ increases, the spectrum seems to concentrates at smaller values of eigenvectors. 
In Appendix~\ref{app:largebeta}, we show that although the number of eigenvectors decreases with $\beta$, it still grows exponentially with $N$, being bounded from below by the integral of the signed density, itself independent of $\beta$.

\section{The largest eigenvalue}
\label{sec:largest}

\subsection{Signed distribution}
\begin{figure}
    \centering
    \includegraphics[width=0.4\textwidth]{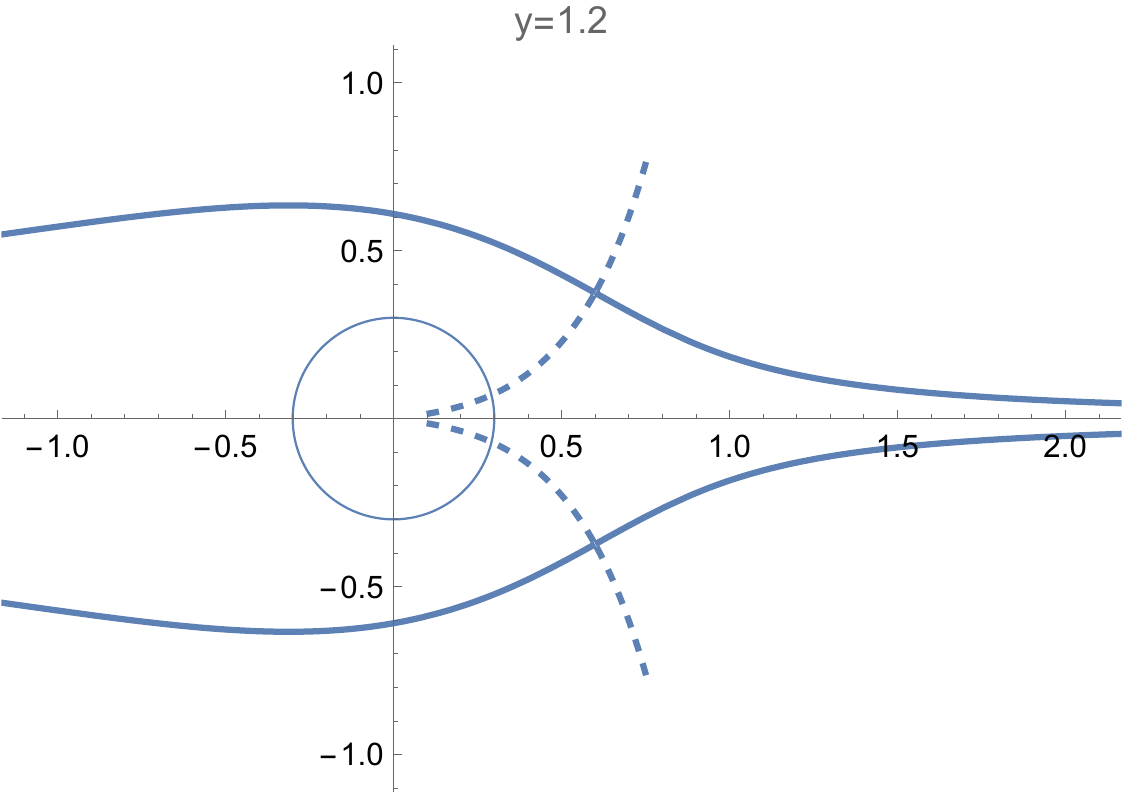}
    \includegraphics[width=0.4\textwidth]{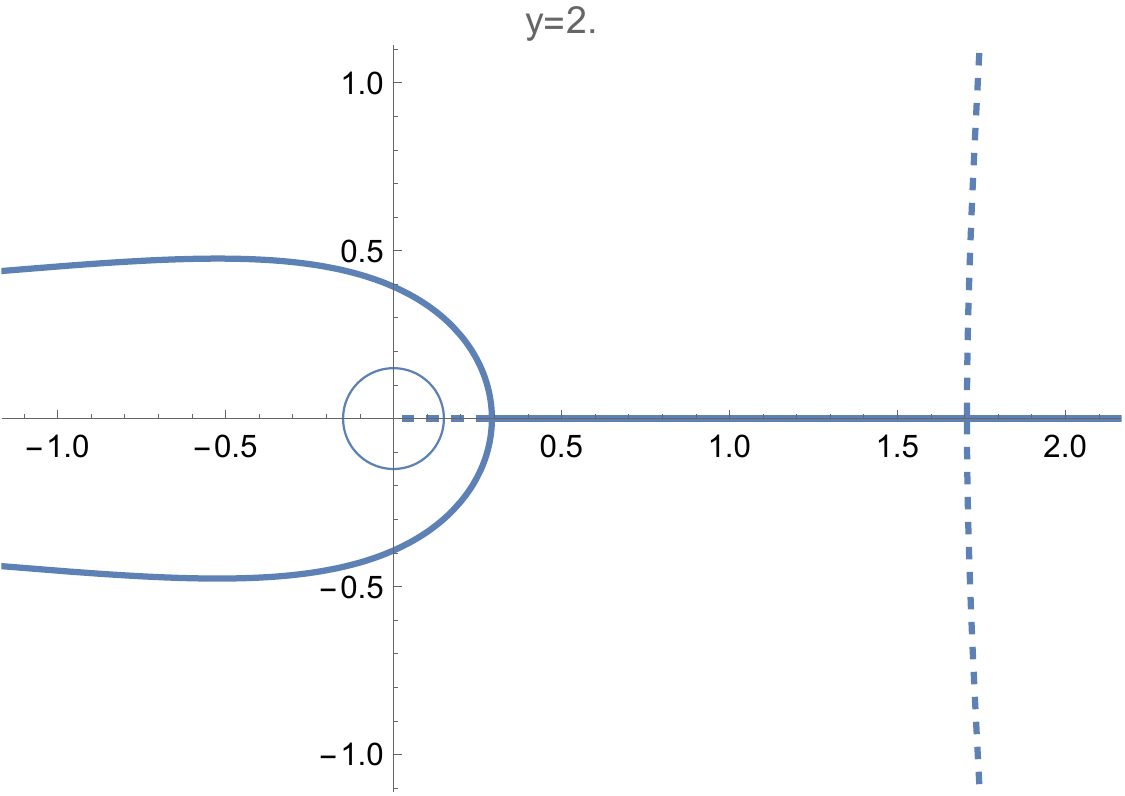}
    \caption{
    On the left, two saddle points and their Lefschetz thimbles (continuous thick line) are relevant, because the two dual Lefschetz thimbles (in broken lines) emerging from them intersect the contour $\mathcal{C}$ (continous thin line), while on the right, only one is relevant.}
    \label{fig:Lefschetz}
\end{figure}

 In order to determine the location of the smallest non-trivial eigenvector $v_{\text{edge}}$, we look at the dominant behaviour in the exponential of the distribution $\rho^{\text{size}}_s(|v|)$. We first analyse $Z_\perp(k)$, following \cite{Kloos:2024hvy}. One can express the confluent hypergeometric function $U$ with Hermite polynomials in the following way,
\begin{gather}
U\left(1-\f{N}{2},\f{3}{2},\frac{3\a k^2}{2\abs{v}^2}\right)=\f{\G(\f{N}{2})\G(\f{N+1}{2})}{\G(N)}\f{1}{\sqrt{\pi}}\f{1}{\sqrt{N} y k} H_{N-1}(\sqrt{N} y k )\,,
\end{gather}
introducing the variable $y = \sqrt{3\a/(2 N \abs{v}^2)}$ that can be seen as a function of $v$. To keep the notation compact, we will sometimes use $y$ and $v$ together in the same formula, but the first has to be understood as function of the latter.
Those polynomials obey the relations
\begin{gather}
H'_{N}(x)=2NH_{N-1}(x)\,,\quad H_N(x)=\frac{N!}{2\pi i}\oint_\cC \dd t \frac{e^{2 t x - t^2}}{t^{N+1}}\,, 
\end{gather}
where the contour $\cC$ winds counterclockwise around the origin. 
The signed density \eq{eq:rhosignfinal} can then be expressed as
\begin{align}
\begs
    \rho^{\text{size}}_s(|v|)
    =&\f{3 \a 2^{N/2}v^{2N-3}}{\sqrt{\pi}}\f{\G(\f{N+1}{2})}{\G(N)}e^{S_0}\left(\frac{\abs{v}^4-4\a \b}{\abs{v}^4+4\a \b}+\frac{8\b \abs{v}^2}{\abs{v}^4+12\a\b}\partial_k\right)\f{1}{\sqrt{N} y} H_{N-1}(\sqrt{N} y k )\bigg|_{k=1}\\
    \sim & \f{3 \a 2^{N/2}v^{2N-3}}{\sqrt{\pi}}\f{\G(\f{N+1}{2})}{\G(N)}e^{S_0}
    \left(\frac{\abs{v}^4-4\a \b}{\abs{v}^4+4\a \b} \frac{N^{-N/2}}{y}+\frac{16 \b \abs{v}^2}{\abs{v}^4+12\a\b} N^{(2-N)/2}\right)\frac{(N-1)!}{2\pi i}\oint \dd \tau e^{N f(y,\tau)}
\ends
\end{align}
in the large $N$ limit, using the variable $\tau=t/\sqrt{N}$ and 
\be
f(y,\tau)=2\tau y - \tau^2 - \log \tau\,.
\ee
The saddle points of $f(y,\tau)$ with respect to $\tau$ are located at 
\be 
\tau_{\pm}=\f{1}{2}(y\pm \sqrt{y^2-2})
\label{eq:taupm}
\ee
with a critical point at $y_c=\sqrt{2}$. 
Since the function $f$ is meromorphic in $\tau$, the contour $\mathcal{C}$ can be deformed to a sum of Lefschetz thimbles. They are defined as the paths of steepest descent from the saddle points of the integrand,\footnote{See for instance 
Section 3 of \cite{Witten:2010cx}
for the steepest descent method in terms of Lefschetz thimbles.}
originating from the different saddle points $\tau_*$ and obeying the flow equation,
\begin{equation}
    \dv{\tau(u)}{u}=-\overline{\pdv{f(y,\tau)}{\tau}}\,, \quad \lim_{u\to-\infty}\tau(u)=\tau_*,
\end{equation}
where $u\in \mathbb{R}$ parametrizes $\mathcal{C}$.
We can identify the saddles that actually contribute in the integral as those whose upward flow, (the so-called dual Lefschetz thimbles), determined by 
\begin{equation}
    \dv{\tau(u)}{u}=\overline{\pdv{f(y,\tau)}{\tau}}\,, \quad \lim_{u\to-\infty}\tau(u)=\tau_*\,,
\end{equation}
intersect the original contour $\cC$. 
As in \cite{Kloos:2024hvy}, only the $\tau_-$ saddle point contributes in the $y>y_c$ region, that corresponds to the small $v$ region, while both $\tau_\pm$ are relevant in the region $y<y_c$, see Fig~\ref{fig:Lefschetz}.

In order to track the displacement of the edge of the density, we gather the leading factors in $N$, we have for $y>y_c$
\begin{align}
\begs
    \rho^{\text{size}}_s(|v|)\sim 
    &\exp(N\left[f(y,\tau_-)-\frac{1}{2}-\frac{1}{2}\log (1+\f{16 \bbeta y^4}{3\a}) -\frac{6\a^2 y^2}{9\a^2+16 \a\bbeta y^4}\right]),
\ends
\end{align}
where we defined $\bbeta=\beta N^2$. Taking also into account the contribution of the second saddle point for $y<y_c$ (in which case the two saddles $\tau_\pm$ give complex conjugate contributions due to the square root in \eqref{eq:taupm}), \footnote{In more detail, the sum of the two contributions is $e^{Nf(y,\tau_-)-i \theta_0}+e^{N f(y,\tau_+)+i \theta_0}= 2 \cos( N{\rm Im}f(y,\tau_-)-\theta_0) e^{N {\rm Re} f(y,\tau_-)}$,
where $\theta_0$ is a phase shift from a sub-leading order. 
Then we can ignore the oscillatory cosine factor, since it is an order one quantity.}
\begin{align}
\begs
    h(y)&=\lim_{N\to \infty}\frac{1}{N}\log \rho^{\text{size}}_s(y) \\
    &= \Re f(y,\tau_-)-\frac{1}{2}-\frac{1}{2}\log (1+\f{16 \bbeta y^4}{3\a}) -\frac{6\a^2 y^2}{9\a^2+16 \a\bbeta y^4}.
    \label{eq:largeNrhoSigned}
\ends
\end{align}

\begin{figure}
    \centering
    \includegraphics[width=0.4\textwidth]{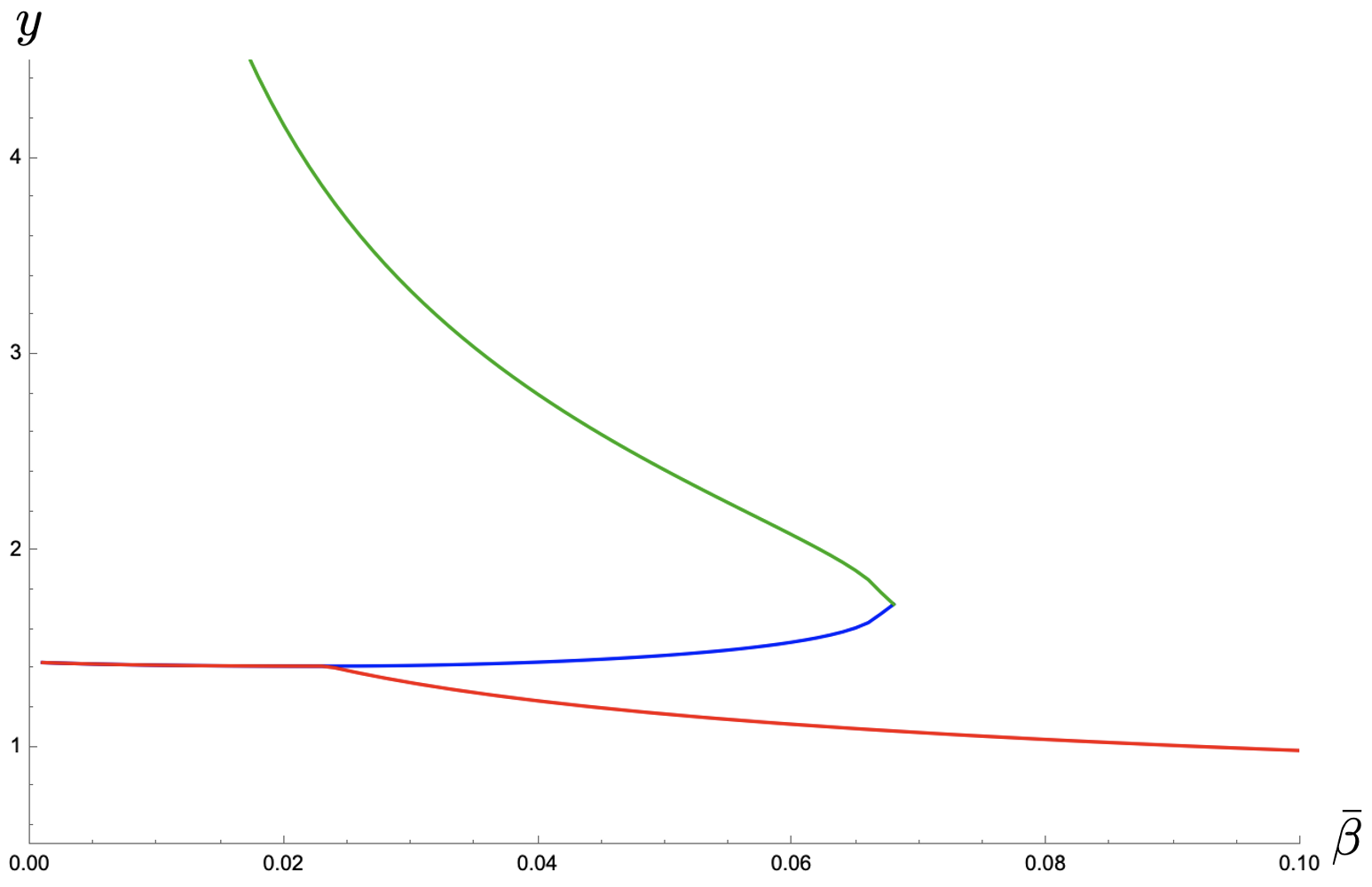}
    \caption{The different real roots of the asymptotics of the signed density as a function of the rescaled noise variance $\bbeta$ ($\alpha=1/2$).}
    \label{fig:rootvsBB}
\end{figure}

\begin{figure}
    \centering
    \includegraphics[width=0.3\textwidth]{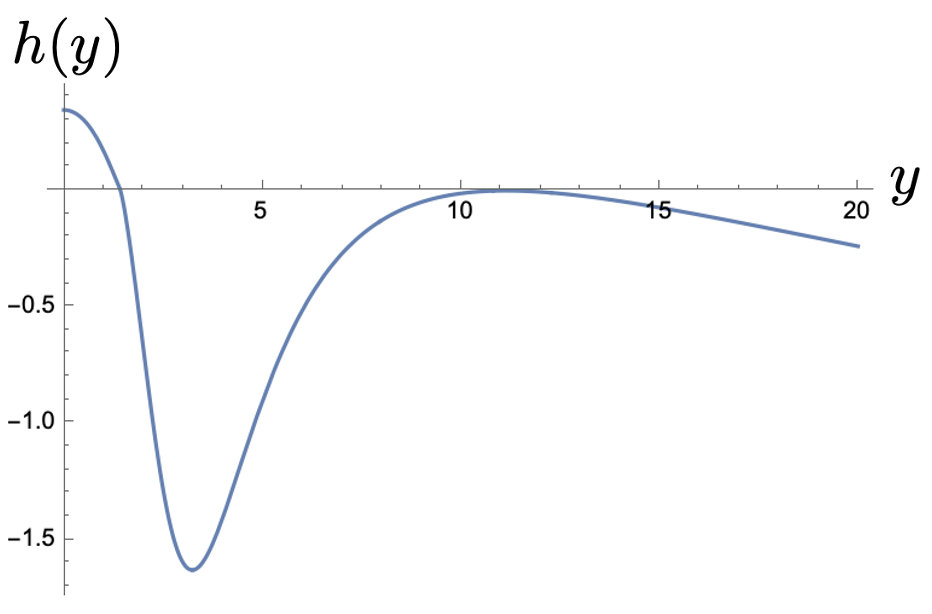}
    \includegraphics[width=0.3\textwidth]{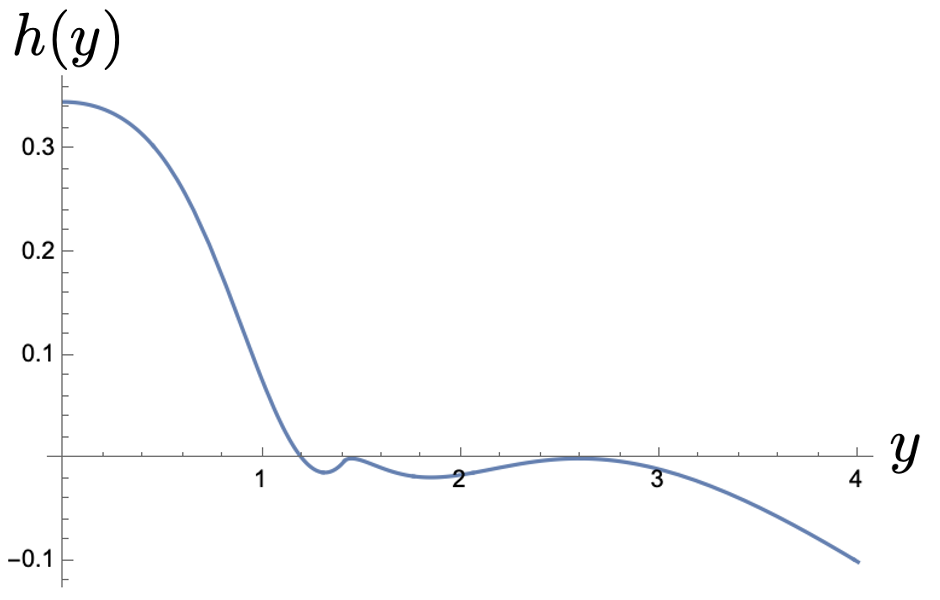}
    \includegraphics[width=0.3\textwidth]{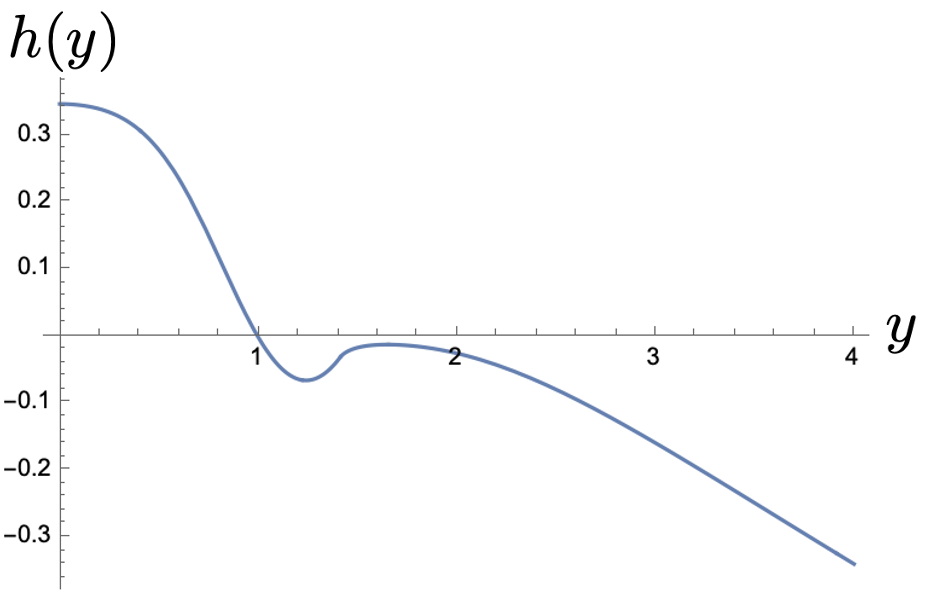}
    \caption{The asymptotics of the logarithm of the signed density as $\bbeta$ increases, from left to right $\bbeta=0.003,0.045,0.095$ ($\alpha=1/2$).}
    \label{fig:logRho}
\end{figure}

We observe three different regimes in the roots of the exponent of distribution $h(y)=0$, represented on Fig~\ref{fig:rootvsBB}, and the corresponding plots of $h(y)$ on Fig~\ref{fig:logRho}, for $\bbeta=0.003,0.045,0.095$ with $\alpha=1/2$.
We conjecture that the Fig~\ref{fig:rootvsBB} gives the limiting behaviour of the largest eigenvalue, and we show numerical support for this behaviour in the App.~\ref{appendix:interpretation} by tracking down the evolution of the first six eigenvalues, as the standard deviation of the noise increases.

There is a first branching point at the location of the critical point $y_c=\sqrt{2}$, where $\bbeta_1$ solves
\begin{equation}
\label{eq:bbetaCusp}
    0 = 1-\frac{12\alpha}{9 \alpha + 64 \bbeta_1} +\frac{1}{2} \log 2 -\frac{1}{2}\log \left(1+\frac{64 \bbeta_1}{3\alpha}\right)\,, \quad \bbeta_1=3\a/64\,,
\end{equation}
giving for $\alpha=1/2$, $\bbeta_1= 3/128\approx 0.023$. 
An expansion near the critical point indicates the following behaviour for $\bbeta<\bbeta_1$
\begin{equation}
    \abs{y-y_c}\sim  \frac{128\sqrt{2}}{9}\abs{\bbeta-\bbeta_1}^2\,,\quad y>y_c\,,
\end{equation}
while on the other side $\bbeta>\bbeta_1$, we have two different exponents
\begin{equation}
    \abs{y-y_c}\sim 
    \begin{cases}
        \frac{512\sqrt{2}}{9}\abs{\bbeta-\bbeta_1}^2\,,& y>y_c\,,\\
        \frac{2^{7/4}}{\sqrt{3}}\abs{\bbeta-\bbeta_1}\,,& y<y_c\,.
    \end{cases}
\end{equation}

The second branching point is located where the two largest real eigenvalues, corresponding to the two maxima 
on the right in the middle picture of Fig.~\ref{fig:logRho}, merge and become complex.  
Here the locations of the two maxima can be obtained by solving the extremal condition, 
\begin{equation}
\frac{\partial h(y)}{\partial y}=0.
\label{eq:hyy}
\end{equation}
The solution of this extremal condition satisfying $y>\sqrt{2}$ and $\bbeta>0$ is given by
\begin{equation}
\bbeta=\frac{3 \alpha (-3 + 2 y^2 + 2 y \sqrt{-2 + y^2})}{16 y^4}.
\label{eq:bbeta_2}  
\end{equation}
Then it is straightforward to see that the $\bbeta$ in \eqref{eq:bbeta_2}
takes the maximum value $\bbeta_2$ at $y=\sqrt{3}$. In other words,
for $\bbeta>\bbeta_2$, there are no solutions to \eqref{eq:hyy} satisfying $y>\sqrt{2}$.
For $\alpha=1/2$, $\bbeta_2=(3 + 2 \sqrt{3})/96\approx 0.067334$. 
One can also check that $h(y)=0$ for $\bbeta=\bar\beta_2,\ y=\sqrt{3}$, implying that the two maxima indeed satisfy $h(y)=0$ as well.

\subsection{Genuine distribution}
The leading order genuine density is given by 
\begin{align}
\begin{split}
    \rho^{\text{size}}(|v|)\sim&\frac{2 \cdot 3^{(N-1)/2}\a^{N/2}}{\Gamma(N/2)}\frac{\abs{v}^{N-1}}{(\abs{v}^4+4\a\b)^{1/2}(\abs{v}^4+12\a\b)^{(N-1)/2}}\exp(-\frac{\a \abs{v}^2}{\abs{v}^4+4\a\b})\\
    &\exp(\frac{N-1}{4}\Re\left(2+\log 16 +\frac{1-\sqrt{1-4 x}}{x}-4\log (1-\sqrt{1-4 x})+4 \log x\right) - (N-1))\,,
\end{split}
\end{align}
with $x=(N-1)|v|^2/3\alpha$ and in the second line, the approximation of the large $N$ field theory partition function was taken, subtracting $S_{eff}(0)=N-1$ to ensure that it is appropriately normalized. \footnote{We also multiplied by a factor $\pi^N$ to account for the two scalar fields $\phi$ and $\sigma$.}
The large $N$ behaviour of the genuine distribution is then
\begin{align}
\begin{split}
    \log \rho^{\text{size}}(x)\sim &\frac{N}{4}\Re\bigg(-\frac{12 \a x}{4\bbeta+9 \alpha x^2}+\log 16 +\frac{1-\sqrt{1-4x}}{x}-4 \log (1-\sqrt{1-4x})+4 \log x\\
    &+2\log 9\alpha x-2\log (3 \alpha (4 \bbeta+3 \alpha x^2))\bigg)\,,
\end{split}
\end{align}
which is the same as eq.~\eqref{eq:largeNrhoSigned}.

\section{Conclusion}
\label{sec:conclusion}

To summarize, extending the series of works \cite{sasakuraSDE,Sasakura_2023,sasakura2023real,sasakura2024signed,Kloos:2024hvy}, we have obtained the exact signed distribution and the large $N$ genuine distribution of the Z-eigenvalues of Gaussian real random fully-symmetric tensors of order 3, in the presence of a Gaussian noise, using respectively, a quartic fermionic theory and an interacting bosonic and fermionic theory, focusing on the location of the largest eigenvalues. Generally, the noise concentrated the spectrum at smaller eigenvectors as it grew. We observed that the edge of the two distributions (signed and genuine) obeyed the same equation, and we found two critical points as the variance of the noise increased: the first, corresponding to the edge of the semi-circle of the random matrix integral \cite{auffinger2013random}, occurred at the emergence of an outlier (the second largest eigenvalue) and the second happened when the (trivial) largest eigenvalue merged with the outlier. At the first transition point, the behavior of the location of the edge of the distribution as a function of the size of the Gaussian
noise changed in a discrete manner that we determined.
We have supported our analytical expressions with numerical simulations.

Some interesting directions to pursue are, the corresponding problem for higher rank tensors or in the line of \cite{sasakura2024signed,dartois2024injective} the associated largest eigenvalue of complex tensors, that is important for measures of entanglement entropy of random quantum states, tensors with more general symmetries or with correlated entries, in order to investigate the existence of universality classes distinct from those of random matrices. 

Another intriguing problem stems from viewing the random tensor as the adjacency ``matrix” of a hypergraph \cite{li2024enhancing}. Understanding the leading eigenvalues of the tensor and their correlations can teach us about the asymptotic stability of the system on such random networks, analogous to the May-Wigner instabilities (e.g. \cite{fyodorov2016nonlinear} for a different way to incorporate non-linearities) or to turbulence \cite{dartois2020melonic}. 
In Appendix~\ref{app:newScaling},  normalizing the noise differently, we show numerical results that indicate that the number of critical points drops to 1, after a certain threshold. A second plot shows how the overlap between corresponding eigenvectors of the same tensor, with and without deviation, drops as the variance increases, reaching zero around $\bbeta_1$. We leave a more thorough exploration of those points to a future study.
Hopefully, one could return from the knowledge of relevant eigenvectors to the study of simplicial complexes built from tensors.

More broadly, large $N$ corrections, fluctuations of the edge of the tensor eigenspectrum, second and higher order moments of the distribution in the spirit of \cite{subag2017complexity,ros2019complex}, and a complete description of the tensor pseudospectrum are an important program that we leave for the future.

\section*{Aknowledgements}
We are grateful for the help and support provided by the Scientific Computing
and Data Analysis section of Core Facilities at OIST that allowed part of the simulations. ND acknowledges the organizers of the YITP-ExU long-term workshop Quantum Information, Quantum Matter and Quantum Gravity (QIMG2023), Yukawa Institute for Theoretical Physics, Kyoto University, during which part of this work was accomplished. We would also like to thank R.~Toriumi for the discussions at the initial stage of this work. 
NS is supported in part by JSPS KAKENHI Grant No.19K03825 and the Kyoto University Foundation, and ND is supported in part by JSPS KAKENHI Grant No.21K20340. The authors acknowledge support of the Institut Henri Poincaré (UAR 839 CNRS-Sorbonne Université), and LabEx CARMIN (ANR-10-LABX-59-01).

\appendix
\section{Numerical simulations}
\label{appendix:num}
Let us explain the setting of our numerical simulations. Similarly to \cite{Sasakura:2022zwc,sasakuraSDE,Sasakura_2023}, we generate a random tensor $\s_{ijk}
$ with components normally distributed as a centered Gaussian of unit variance $\mathcal{N}(0,1)$, and multiply with the appropriate combinatorial factor that accounts for the symmetries of the indices: 
\begin{gather}
c_{ijk}=\f{\s_{ijk}}{\sqrt{d(i,j,k)}},\\
d(i,j,k)=
\begin{cases}
1,\quad i=j=k,\\
3,\quad  i= j\neq k, \text{or }  i\neq j= k, \text{or } i= k\neq j, \\
6, \quad i\neq j\neq k \neq i,
\end{cases}
\end{gather}
ensuring that the measure arises from the norm 
\be 
|c|^2=\sum_{1\leq a,b,c\leq N}c_{abc}^2=\sum_{1\leq i\leq j\leq k\leq N}d(i,j,k)c_{ijk}^2.
\ee
Additionally, we generate a random Gaussian vector $\{\nu_i\}_{1\leq i \leq N}$ of variance $2\b$ to account for the width of the Lagrange multiplier $\l$. We solve the eigenvector equation
\be
v_i=c_{ijk}v_jv_k+\nu_i
\ee
and keep the real solutions, storing their absolute value $\abs{v}$ and the sign of the determinant of the Jacobian matrix $M_{ab}$ for the signed case. The distributions of the absolute value $\abs{v}$ without and with the weight of the sign of the determinant of $M_{ab}$ are represented above in Fig.~\ref{fig:signed} and Fig~\ref{fig:numN10}. We have gathered our Mathematica notebooks at \url{https://github.com/dlprtn/EdgeRandomTensors}.

\section{Finite $N$ distribution and interpretation of the solutions of $h(y)=0$}
\label{appendix:interpretation}

We first elaborate on how a finite $N$ expression for the eigenvector distribution can be obtained from \cite{auffinger2013random}. We start from a rather trivial expression,
\be
\label{eq:rhobeta1}
\rho^{\rm size}(|v|,\alpha,\beta)=\frac{\rho^{\rm size}(|v|,\alpha,\beta)}{\rho^{\rm size}(|v|,\alpha,\beta=0)} \rho^{\rm size}(|v|,\alpha,\beta=0),
\ee
and take the advantage that the exact expression of $\rho^{\rm size}(|v|,\alpha=1/2,\beta=0)$ 
for arbitrary $N$ is given in \cite{auffinger2013random}. 
The relation between their variable $u$ and our $\abs{v}$ is given by
\begin{equation}
\frac{1}{|v|}=-\sqrt{N} u,
\label{eq:relation} 
\end{equation}
which can be derived by comparing the stationary condition of the Hamiltonian (energy) in \cite{auffinger2013random} and our eigenvector
equation (see an appendix of \cite{Sasakura_2023} for an explicit derivation). 
Our $\rho^{\rm size} (|v|,\alpha=1/2,\beta=0)$ is equal to the derivative of 
(2.9) of\cite{auffinger2013random} with respect to $|v|$, 
considering that they are the number density 
and the number, respectively. Then, by using (7.4)-(7.8) of \cite{auffinger2013random},
we obtain the exact expression of $\rho^{\rm size} (|v|,\alpha=1/2,\beta=0)$.\footnote{More explicitly, 
$\rho^{\rm size}(|v|,\alpha=1/2,\beta=0)=2\sqrt{N}|v|^{-2}(p-1)^{(N-1)/2} \exp (-(p-2)/(4(p-1) |v|^2)) 
\rho_N(\tilde x)$, where $\rho_N(\cdot)$ is that of \cite{auffinger2013random},
$\tilde x=-\sqrt{p/(2(p-1))}/(\sqrt{N} |v|)$, and $p=3$.}

Next let us estimate the first factor of \eqref{eq:rhobeta1}. 
From \eqref{eq:FT} and using the approximation \eq{eq:zftproduct}, 
we find 
\be
\label{eq:rhoapp}
\frac{\rho^{\rm size}(|v|,\alpha,\beta)}{\rho^{\rm size}(|v|,\alpha,\beta=0)}\sim \left(\frac{|v|^4}{|v|^4+4 \alpha \beta}\right)^{\frac{1}{2}}
\left(\frac{|v|^4}{|v|^4+12 \alpha \beta}\right)^{\frac{N-1}{2}}\exp \left( -\frac{\alpha |v|^2}{|v|^4+4 \alpha \beta} 
+\frac{\alpha}{|v|^2} \right) \frac{Z_{FT}^\parallel }{Z_{FT}^\parallel (\beta=0)}, 
\ee
where $Z_{FT}^\perp$ cancels out, because
it is independent of $\beta$.
Note that $Z_{FT}^\parallel$ can approximately be expressed by \eqref{eq:zpara} with $Q_{ij}$ as done in the main text.

Let us argue now that the roots of $h(y)=0$,
which are not the boundary of the $h(y)>0$ region, correspond to single eigenvectors which are isolated from the main part of the distributions. Having to deal with single eigenvectors, we could not rely on the large $N$ expressions. As shown in Fig.~\ref{fig:rootvsBB}, 
the number of solutions of $h(y)=0$ changes depending on the value of $\bar \beta$:
two solutions for $\bar \beta < \bar \beta_1$, three for $\bar \beta_1 < \bar \beta  <\bar  \beta_2$, and 
one for $\bar \beta_2 < \bar \beta$. Since $h(y)>0$ in the region $y<y_{\rm smallest}$, where $y_{\rm smallest}$ 
denotes the smallest root of  $h(y)=0$, it is clear that $y_{\rm smallest}$ is the edge of the distribution
in the large-$N$ limit. On the other 
hand, the other roots of $y$ are not edges of any regions of $h(y)>0$, 
and the interpretation of these solutions was not so clear.

By generalizing the arguments of \cite{Kloos:2024hvy}, 
the locations of the eigenvectors can be estimated by solving
\be
\label{eq:rhosizecond}
\int_0^{|v_{n}|} \dd z\, \rho^{\rm size} (z,\alpha,\beta)=n-\frac{1}{2},
\ee
where $\abs{v_n}$ denotes the estimation of the absolute value of the $n$-th smallest eigenvector,
and we can use the above result for $\rho^{\rm size} (z,\alpha,\beta)$ for $\alpha=1/2$. 

\begin{figure}
    \centering
    \includegraphics[width=0.4\textwidth]{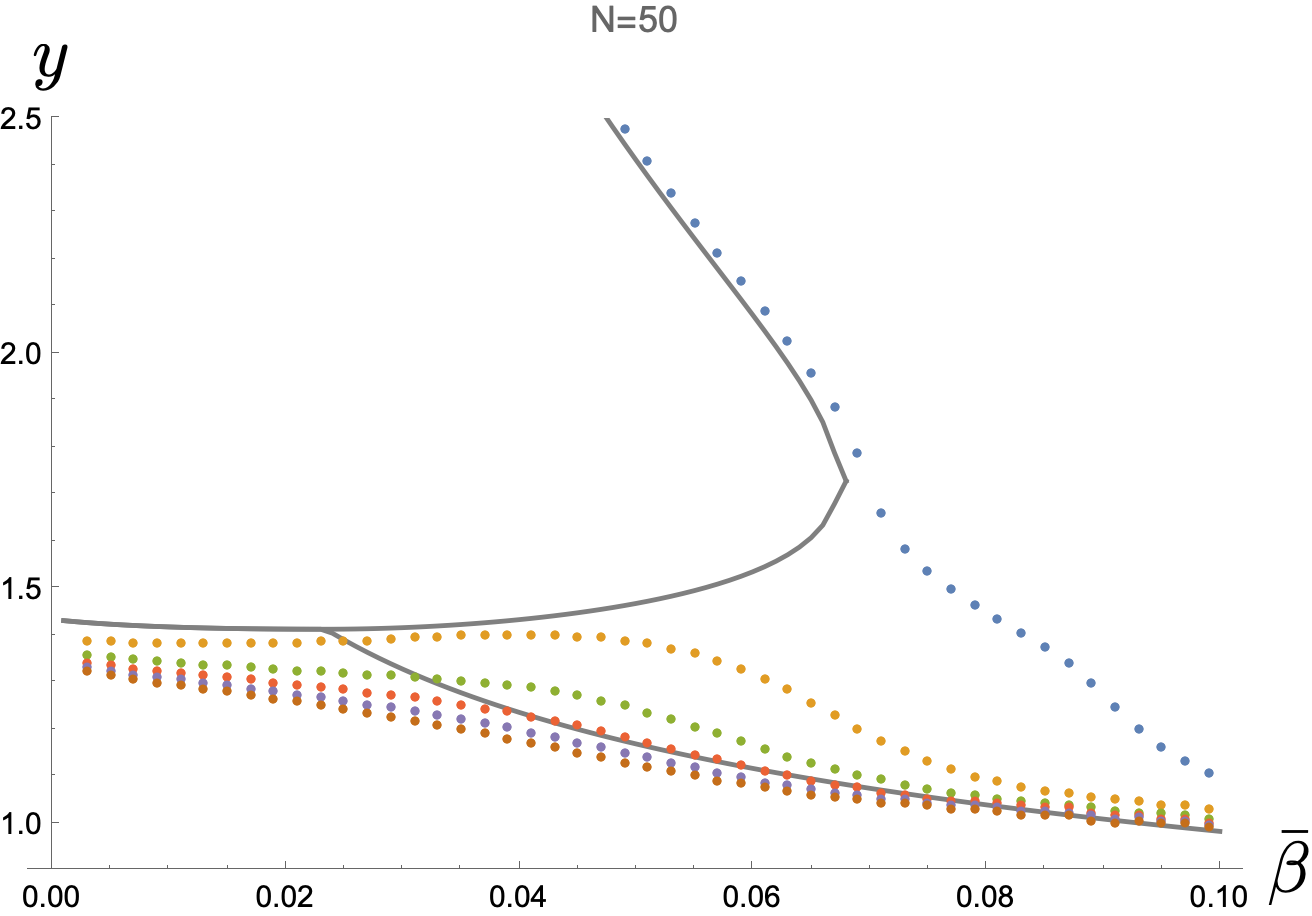}
    \caption{The solutions to \eqref{eq:rhosizecond} up to $n=6$ are plotted in $y$ against $\bar \beta$ for $N=50$. The lines of Fig.~\ref{fig:rootvsBB} are also drawn on the figure for comparison.}
    \label{fig:n45}
\end{figure}

In Fig.~\ref{fig:n45}, the locations 
of the smallest eigenvectors up to $n=6$ are plotted in $y$ against $\bar \beta$ for $N=50$ by solving \eqref{eq:rhosizecond}.
The value $N=50$ still seems too small for a definite conclusion,
but, comparing with Fig.~\ref{fig:rootvsBB}, the behavior of the 
smallest two eigenvectors seems consistent with what the roots of $h(y)=0$ imply:
the first two smallest eigenvectors can be identified with the isolated roots found for
$\bar \beta_1 < \bar \beta<\bar \beta_2$, and they ``disappear" after $\bar \beta=\bar \beta_2$. 
(More precisely, in the region $\bbeta_2<\bbeta$, 
the smallest eigenvector is succeeded by the third one.)

\section{A comment on the large $\beta$ limit}
\label{app:largebeta}
\begin{figure}
\begin{center}
\includegraphics[width=7cm]{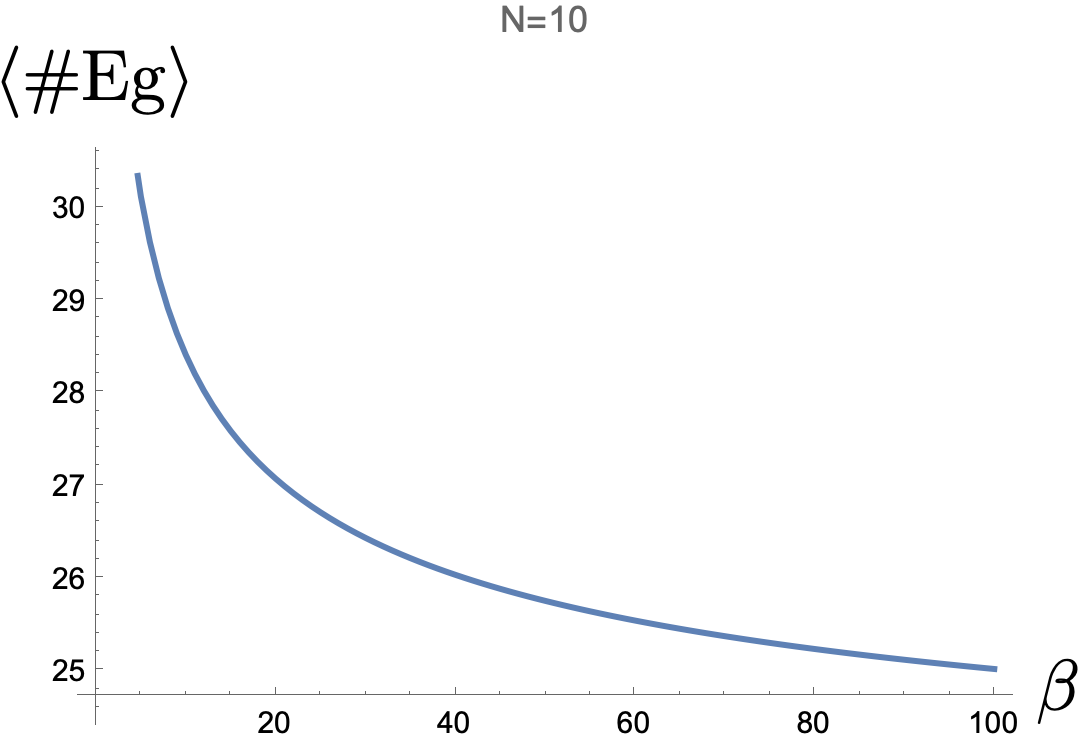}
\caption{The behavior of the mean number of eigenvectors against $\beta$ for $N=10$. 
The number decreases, as the noise size $\beta$ becomes larger.}
\label{fig:egnum}
\end{center}
\end{figure}

As is shown in an example on Fig.~\ref{fig:egnum}, 
the effect of taking larger $\beta$ generally reduces the 
total number of eigenvectors. The question we want to discuss here is whether the number vanishes 
in the large $\beta$ limit. We will show that there remain an exponentially large number of them. 

Since the difference between the genuine and signed distributions is to take the absolute value of the 
determinant or not, the mean number of eigenvectors is bounded from below by the signed distribution as 
\be
\langle \# {\rm Eigenvectors}\rangle=\int_{\mathbb{R}^N} \dd v \, \rho(v) 
\geq \left | \int_{\mathbb{R}^N} \dd v \, \rho_s(v) \right|. 
\label{eq:bound}
\ee
In fact, $\int_{\mathbb{R}^N} \dd v \, \rho_s(v)$ is constant against $\beta$, as will be shown below.
Therefore, if it is non-zero,  $\langle \# {\rm Eigenvectors}\rangle$ is bounded 
from below by a positive number even for $\beta\rightarrow\infty$. 

Let us prove the constancy of  $\int_{\mathbb{R}^N} \dd v\, \rho_s(v)$ against $\beta$. We first observe that, since we integrate
over $v$, $v$ can be regarded as a dynamical variable of the system. We introduce
the following BRST symmetry $Q_{\rm BRST}$ for the variables $v,\lambda, \bar \psi, \psi$:
\be
\qb v_a=\bar \psi_a,\ \qb \lambda_a =0,\ \qb \bar \psi_a=0,\ \qb \psi_a=i \lambda_a,
\ee
which indeed satisfies $Q_{BRST}^2=0$. Then we observe 
\be
\qb \left( \left(v_a-C_{abc}v_b v_c \right)\psi_a \right)
=i \lambda_a (v_a-C_{abc}v_b v_c)+\bar \psi_a (\delta_{ab}-2 C_{abc}v_c ) \psi_b,
\ee
which agrees with the latter two terms of \eqref{eq:signaction}. This implies that those terms form a
BRST invariant action. In addition, we can further observe
\[
i \qb\left( \lambda_a \psi_a \right)= -\lambda^2,
\]
which means that the second term with $\beta$ in \eqref{eq:signaction} is in fact BRST exact. 
By a standard argument, the partition function
\be
\int \dd v \dd\lambda \dd\bar \psi \dd \psi\, e^{-\beta \lambda^2 + i \lambda_a (v_a-C_{abc}v_b v_c)+\bar \psi_a (\delta_{ab}-2 C_{abc}v_c ) \psi_b}
\ee
is invariant under the change of $\beta$, which proves the invariance of the integral of the signed distribution $\int_{\mathbb{R}^N} \dd v \rho_s(v)$
under the change of $\beta$. 

It is not difficult to compute the explicit value of the bound \eqref{eq:bound}. 
The integration of $\rho_s(v)$ over $v$ can most easily be computed for $\beta=0$.
Putting $\beta=0$ into \eqref{eq:rhosignfinal}, we obtain  
\begin{align}
\begin{split}
\label{eq:rhosint}
\int_0^\infty \dd |v|\, \rho_s^\text{size}(|v|,\beta=0) 
&= \frac{\sqrt{3}\alpha 2^{N/2} }{\Gamma(N/2)} \int_0^\infty \dd |v| \ |v|^{-3} e^{-\alpha/|v|^2}
U\left(1-N/2,3/2,3 \alpha/(2 |v|^2)\right)
\cr
&=\left\{ 
\begin{array}{ll}
-1+\frac{(-1)^{N/2} 2^{-N/2+1} \Gamma(N)}{\sqrt{3} \Gamma(N/2)\Gamma(N/2+1)} 
{}_2 F_1(1/2, 1; N/2+1; 2/3) & \hbox{for even }N, \\
-1 & \hbox{for odd }N,
\end{array}
\right.
\end{split}
\end{align}
where ${}_2 F_1(\cdot,\cdot;\cdot;\cdot)$ is the hypergeometric function.
In fact, $\rho_s^\text{size}(|v|,\beta=0)$ is missing the trivial solution $v=0$, as can be seen
in its expression. Since the $v=0$ solution always exists, and ${\rm det} M=1$ for the solution, 
one should add 1 to the result of \eqref{eq:rhosint}, which cancels the first terms of 
\eqref{eq:rhosint}. \footnote{The main reason why we have to count the trivial solution $v=0$
for $\beta=0$ is that the trivial solution is moved to a $v\neq 0$ solution for $\beta>0$, and therefore
is counted in the integral $\int_0^\infty d|v|\, \rho_s(|v|,\beta>0)$.}
Thus we obtain a bound,
\be
\label{eq:boundhyper}
\langle \# {\rm Eigenvectors} \rangle \geq \frac{2^{-N/2+1} \Gamma(N)}{\sqrt{3} \Gamma(N/2)\Gamma(N/2+1)} 
{}_2 F_1(1/2, 1; N/2+1; 2/3) \sim \frac{2^{(N+1)/2}}{\sqrt{3 \pi N}}
\ee
for even $N$, 
where the asymptotic behavior for large $N$ has been estimated.
For odd $N$, we cannot derive a meaningful bound from \eq{eq:bound}, but \eqref{eq:boundhyper} 
could be extended to the odd case too, since $\langle \# {\rm Eigenvectors} \rangle$ is numerically 
a smooth function of $N$, no matter if $N$ is odd or even.  
We see that, 
even for $\beta\rightarrow \infty$, there remain an exponentially large number of eigenvectors. 
 
\section{Numerical experiments}
\label{app:newScaling}
\subsection{Another rescaling for $\beta$}
If one adds the Gaussian deviation $\tilde{\nu}_i$ in the eigenvalue equation \eqref{eq:SphWithRandomField}, analogously to \cite{fyodorov2015high}, we obtain the corresponding eigenvector equation
\be
\label{eq:eigvecdeviationFyod}
\sum_{1\leq i_2\dots i_p\leq N}C_{i_1\dots i_p}v_{i_2}\dots v_{i_p}= v_{i_1}+\tilde{\nu}_{i_1} \abs{v}^{p-1}\,, \quad (1\leq i_1\leq N, v \neq 0, v \in \mathbb{R}^N)\,
\ee
We noticed a crucial difference in the total number of real eigenvectors between the two normalizations \eqref{eq:eivec-def} and \eqref{eq:eigvecdeviationFyod}, represented on the Fig.~\ref{fig:compareNoise} for $p=3$ and $N=10$. The latter choice of scaling seems to suggest a topological trivialization in agreement with \cite{fyodorov2015high}. We leave the confirmation of this observation and the computation of the corresponding effective action for future work.

\subsection{Eigenvector overlap}
\begin{figure}
\begin{center}
\includegraphics[width=7cm]{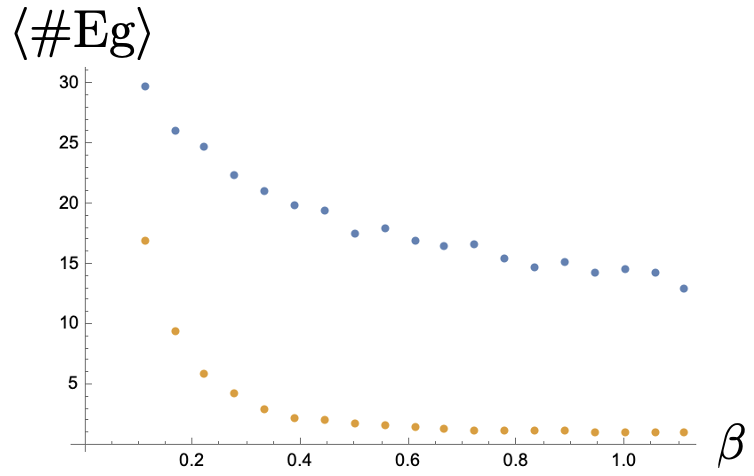}
\caption{Comparison of the mean number of eigenvectors against $\beta$ for $p=3$ and $N=10$ (with 1000 samples per $\beta$). The top curve uses the normalization \eqref{eq:eivec-def}, while the bottom one uses the normalization \eqref{eq:eigvecdeviationFyod}.}
\label{fig:compareNoise}
\end{center}
\end{figure}

Searching for another consequence of the emergence of an outlier, we looked at the inner product between eigenvectors with and without noise, attempting to see how strongly the spectrum was disturbed by the presence of the Gaussian deviation. We represent on Fig.~\ref{fig:compareNoise} the evolution with $\bbeta$ of the average of such inner products for the second and third smaller eigenvectors of the same tensor, with and without deviation. We notice that they approach zero, around the critical point $\bbeta_1$ for the inner product of the third eigenvectors, and seemingly further for the the second eigenvectors. This indicates that beyond the value of $\bbeta_1$, the eigenvectors in the presence of deviation are almost orthogonal to the ones without.
Further study would nonetheless be needed in order to determine the precise location of these points and if they relate to the large-$N$ outlier. 

\begin{figure}
\begin{center}
\includegraphics[width=9cm]{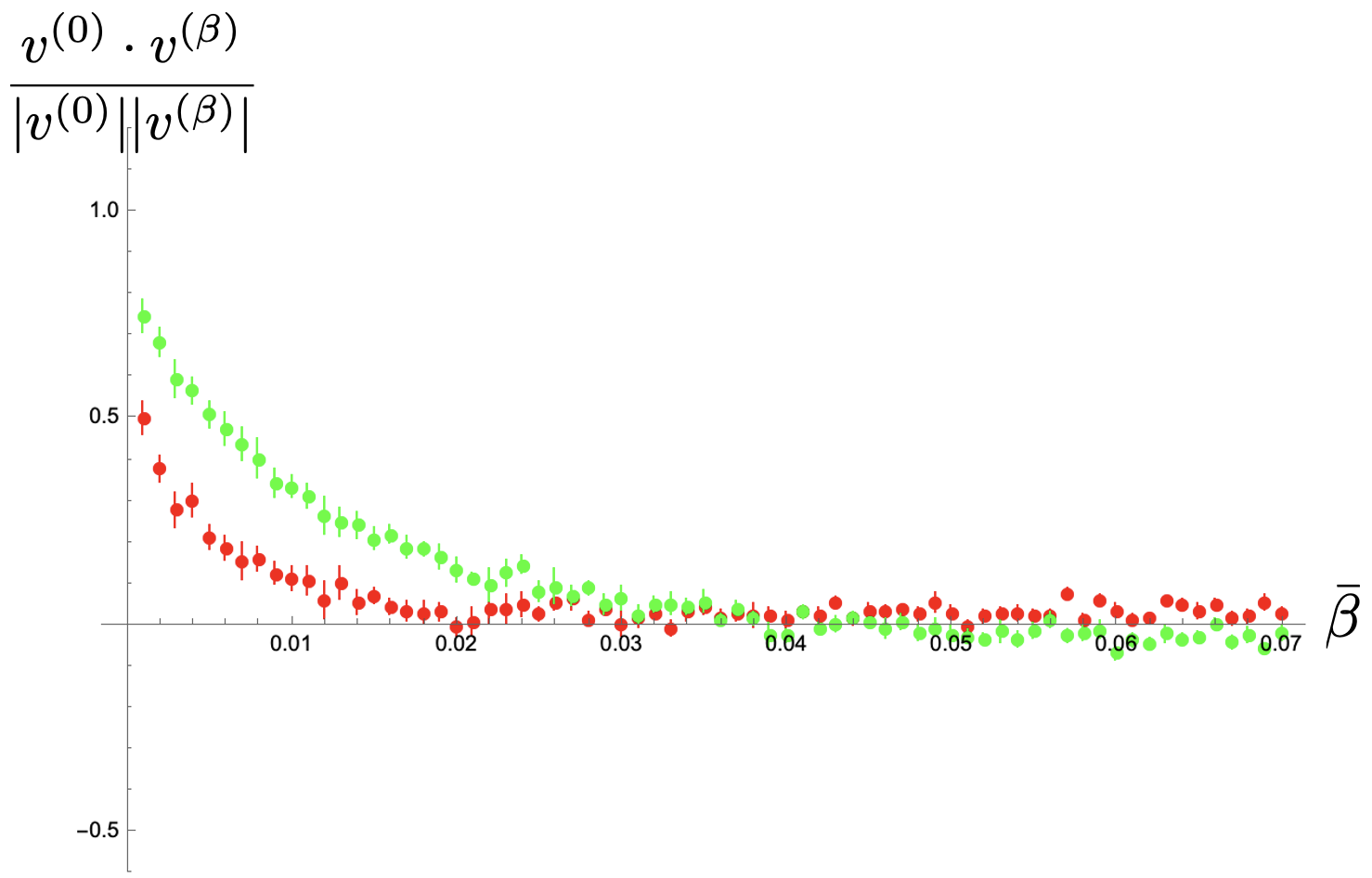}
\caption{We represent against $\bbeta$ the inner products between the corresponding eigenvectors with ($v^{(\beta)}$) and without noise ($v^{(0)}$), for the second (top green curve) and third (bottom red curve) smallest eigenvectors.
The plots are constructed by an extrapolation to $N\rightarrow \infty$ 
from the Monte Carlo data of $4\leq N\leq 12$ with 10000 samples for each $N$ and $\bar \beta$.}
\label{fig:innervvbeta}
\end{center}
\end{figure}

\let\oldbibliography\thebibliography 
\renewcommand{\thebibliography}[1]{\oldbibliography{#1}
\setlength{\itemsep}{-1pt}}
\bibliographystyle{JHEP}
\bibliography{TensorEigenvalues.bib}

\end{document}